\documentclass[trackchanges,twocolumn]{aastex62}
\usepackage{lipsum}
\usepackage{amssymb}
\usepackage{float} \usepackage{listings}
\usepackage{xcolor}
\usepackage{amsmath}
\usepackage{CJK}
\usepackage{listings}

\lstdefinelanguage{SQL}{
    morekeywords={SELECT, FROM, WHERE, AND, AS, LOG10},
    sensitive=false,
    morecomment=[l]{--},
    morestring=[b]',
}

\begin{document}
\makeatletter
\let\frontmatter@title@above=\relax
\makeatother

\newcommand\lsim{\mathrel{\rlap{\lower4pt\hbox{\hskip1pt$\sim$}}
\raise1pt\hbox{$<$}}}
\newcommand\gsim{\mathrel{\rlap{\lower4pt\hbox{\hskip1pt$\sim$}}
\raise1pt\hbox{$>$}}}
\newcommand{\CS}[1]{{\color{red} CS: #1}}

\title{\Large \textbf{A global view of post-interaction white dwarf-main sequence binaries}}

\shorttitle{A global view of post-interaction WDMS binaries}
\shortauthors{Shariat \& El-Badry}

\author[0000-0003-1247-9349]{Cheyanne Shariat}
\affiliation{Department of Astronomy, California Institute of Technology, 1200 East California Boulevard, Pasadena, CA 91125, USA}

\author[0000-0002-6871-1752]{Kareem El-Badry}
\affiliation{Department of Astronomy, California Institute of Technology, 1200 East California Boulevard, Pasadena, CA 91125, USA}

\correspondingauthor{Cheyanne Shariat}
\email{cshariat@caltech.edu}

\begin{abstract}
Common-envelope evolution (CEE) is among the most uncertain phases in binary evolution. To empirically constrain CEE, we construct a uniformly selected sample of eclipsing post--common-envelope binaries (PCEBs). Starting from an unresolved white dwarf--main-sequence (WDMS) candidate sample within 200 pc selected from the {\it Gaia} color-magnitude diagram, we identify 39 detached eclipsing WDMS binaries using ZTF light curves. The binaries contain cool M dwarfs orbiting warm white dwarfs with orbital periods ($P_{\rm orb}$) of $0.1-2$~d. The sample's simple selection function allows us to model observational incompleteness and infer intrinsic properties of the PCEB population. We find an orbital-period distribution consistent with being log-uniform over $0.1-2$ d, contrary to recent reports of a bimodal distribution. The intrinsic companion-mass distribution peaks around $0.25~{\rm M_\odot}$ and declines steeply toward larger masses. The estimated local space density is $7.2\times10^{-5}~{\rm pc^{-3}}$, corresponding to a Galaxy-wide birth rate of $0.01~{\rm yr^{-1}}$. Combining our results with recent {\it Gaia}-based constraints on wider WDMS binaries, we construct an empirical period distribution of post-interaction WDMS binaries spanning $0.1-1000$ d. The emerging period distribution is roughly log-flat (${\rm d}N/{\rm d}\log P_{\rm orb}\propto P_{\rm orb}^0$) at $P_{\rm orb} < 2$ d and log-increasing (${\rm d}N/{\rm d}\log P_{\rm orb}\propto P_{\rm orb}^1$) at $P_{\rm orb} = 100-1000$ d. The $10-100$ d regime is poorly constrained by existing surveys, but a few nearby systems suggest that it is also well-populated. Short-period PCEBs ($P_{\rm orb}\lesssim 2$ d) with M dwarf companions are roughly 2-3 times more common than wide ($P_{\rm orb} = 100-1000$ d) WDMS binaries with FGK companions, which likely formed through stable mass transfer. These results provide direct observational constraints on CEE and an empirical benchmark for binary-population models.
\end{abstract}

\section{Introduction}\label{sec:intro}
Sun-like stars are often born in binary star systems with separations $\lsim10$~au \citep{DM91,Raghavan2010}. As the more massive (primary) star in the binary evolves into a giant, its radius will expand and potentially initiate mass transfer with the companion. 
If this mass transfer is dynamically unstable, the expanding envelope of the primary engulfs the secondary star, leading to common-envelope evolution (CEE). As the secondary spirals through the primary's envelope, orbital energy and angular momentum are drained from the binary and can be used to eject the envelope of the primary star. If enough orbital energy is released to completely expel the primary's envelope, the system emerges as a short-period post--common-envelope binary (PCEB) containing a young (hot) white dwarf (WD) and a main-sequence (MS) star; otherwise, the stars coalesce. 

CEE is therefore a brief but transformative phase that determines whether a binary survives as a close pair or merges \citep{Paczynski76,Ivanova13}. 
Despite decades of study, the physics of common envelope ejection, and thus the final separations and survival rates of binaries undergoing CEE, remain poorly understood \citep{Webbink08,Ivanova13,Ivanova20}.

Compact ($P_{\rm orb}\lsim10$~days) white dwarf--main sequence (WDMS) binaries are the best-observed class of PCEBs. They are (a) relatively common \citep[$10^3-10^4~{\rm kpc^{-3}}$;][]{Schreiber03,RebassaMansergas21,Inight21} and (b) direct progenitors of cataclysmic variables (CVs), ultracompact binaries, highly magnetic WDs, and type~Ia supernovae \citep[e.g.,][]{Warner95,Schreiber03,GarciaBerro12,Parsons13}.
Studying the WDMS PCEB population, whose present-day orbital and stellar properties retain imprints of CEE, is one of the most promising avenues for constraining CEE.

Large spectroscopic surveys such as SDSS identified $\sim100$ WDMS PCEBs through multi-epoch radial-velocity (RV) variability \citep[][]{RebassaMansergas07}. The SDSS PCEB catalog provided one of the first statistical PCEB samples with orbital periods of hours to days \citep{RebassaMansergas07,RebassaMansergas10,RebassaMansergas12}. Extensive work has gone into modeling the SDSS selection function and quantifying the survey’s sensitivity \citep[e.g.,][]{Schreiber10,RebassaMansergas10,GomezMoran11}, enabling population-level inferences about CEE and post--common-envelope binary evolution. As a result, the SDSS sample has played a central role in constraining the demographics of PCEBs and informing theoretical models of CEE \citep[e.g.,][]{Schreiber08,Schreiber10,RebassaMansergas09,GomezMoran09,Schwope09,Zorotovic10,Zorotovic11a,Zorotovic11b,Zorotovic14,RebassaMansergas12b,GomezMoran11,Camacho14,Ablimit16}.

Nonetheless, the SDSS PCEB catalog represents only one spectroscopic survey, and independent constraints from different samples are essential. With precise {\it Gaia} parallaxes now available, WD and companion masses, radii, and space densities can be determined more readily than was possible with the pre-{\it Gaia} SDSS samples.
In parallel, a large population of wide ($P_{\rm orb} \in10-1000$~days) post-interaction WDMS binaries has emerged \citep{ Wonnacott93,Kruse14,Kawahara18, Oomen18, Escorza19, Jorissen19, Masuda19, Hernandez21, Kruckow21,Hernandez22a, Hernandez22b, Shahaf23, Shahaf24, Yamaguchi24b, Yamaguchi24c, Yamaguchi24, Yamaguchi25}: a regime that the SDSS PCEB sample was not sensitive to.
Together, these developments motivate a fresh reassessment of PCEBs using complementary observational strategies to obtain a more complete and coherent picture of the post-interaction WDMS binary population.

A natural next step beyond SDSS is to identify PCEBs photometrically using wide-field time-domain surveys. \citet{Shani25} recently carried out such an effort by searching for WDMS eclipsing binaries in TESS light curves. Their study demonstrates the promise of eclipse-based selection for PCEB studies, but several practical limitations complicate their results.
First, their search was performed using unresolved WDMS binary candidates from \citet{Li25}, a catalog constructed with a neural-network classifier trained on low-resolution {\it Gaia} XP spectra. While this catalog is valuable for discovery work, its selection function is difficult to quantify, making it challenging to assess completeness and biases of the discovered WDMS population.
In addition, the large TESS pixels ($21''$) make contamination from blended sources a potential concern, and a significant fraction of their objects classified as ``eclipsing" do not exhibit WDMS eclipse morphologies (see Section \ref{subsec:TESS}). This contamination propagates into the inferred PCEB fraction and period distribution, making population modeling difficult.

These challenges motivate a complementary approach using a more well-defined parent sample and light curves from the Zwicky Transient Facility \citep[ZTF;][]{Bellm19,Graham19,Masci19}. ZTF's higher spatial resolution and decade-long baseline make it well-suited for constructing homogeneous samples. Moreover, ZTF has already proven effective at identifying eclipsing PCEBs \citep[e.g.,][]{Keller22,Brown23,Li24}, but past works have not yet translated discoveries into population-level constraints.

In this work, we first construct a parent sample of unresolved WDMS binaries directly from the {\it Gaia} color--magnitude diagram (CMD) with simple selection cuts (Section \ref{subsec:parent_sample}). From this parent sample, we search for eclipsing binaries in ZTF (Section \ref{subsec:EB_search}); eclipsing PCEBs provide particularly clean constraints on CEE outcomes, since their detectability is straightforward to model.
Next, we leverage the clean selection function to quantify our sample's completeness (Section \ref{subsec:completeness}), correct for selection biases, and derive empirical constraints on the {\it intrinsic} population demographics of PCEBs (Section \ref{sec:results}). This includes the intrinsic PCEB period distribution (Section \ref{subsec:period_dist}), companion mass distribution (Section \ref{subsec:mass_dist}), fraction (Section \ref{subsec:fraction_pceb}), space density (Section \ref{subsec:space_density}), and birth rate (Section \ref{subsec:birth_rate}).
Section \ref{sec:discussion} explores the role of magnetic braking and compares our results to previous work.
Section \ref{sec:comp_to_wideWDMS} discusses our findings within the broader population of post--mass-transfer WDMS binaries.
We summarize our conclusions in Section \ref{sec:conclusions}. Supplementary data, figures, and modeling details are provided in Appendix \ref{app:sensitivity}, \ref{app:all_LCs}, \ref{app:SED_fitting}, and \ref{app:MB_equations}.

\section{Methodology}\label{sec:methods}

Our goal is to derive a sample of WDMS eclipsing binaries (EBs) while maintaining a well-defined selection function throughout. To do this, we first construct a volume-limited sample of unresolved WDMS candidates from the \textit{Gaia} CMD. From this parent sample, we search for EBs in ZTF photometry and then characterize the completeness of our eclipse-recovery pipeline.

\subsection{Parent sample}\label{subsec:parent_sample}

We select unresolved WDMS sources from \textit{Gaia} by identifying stars located between the main sequence and WD cooling sequence in the CMD. The search is restricted to sources within $200$~pc ($\varpi>5$~mas) with high-quality parallaxes ($\varpi/\sigma_\varpi \ge 10$) and reliable color measurements. We also restrict to sources observable by ZTF: $\delta>-28^\circ$ and $G \gtrsim 12.5$~mag, roughly corresponding to the southern pointing limit and saturation limit of ZTF, respectively. This initial query returns $779{,}142$ sources within $200$~pc.
The ADQL query used to construct the initial sample is:

\begin{lstlisting}[language=SQL]
select * from gaiaedr3.gaia_source
where parallax > 5
  and parallax_over_error > 10
  and phot_g_mean_mag > 12.5
  and phot_rp_mean_flux_over_error > 25
  and phot_bp_mean_flux_over_error > 25
  and dec > -28
\end{lstlisting}

Given that the CMD region between the MS and WD cooling track is intrinsically sparsely populated, even a small fraction of MS stars scattered into this region can easily dominate the sample. To suppress such contamination, we apply astrometric and blending-quality cuts using the reliability metrics of \citet{Rybizki22}. Specifically, we require
(a)  ${\tt fidelity\_v2}>0.9$ to remove sources with spurious or non-single-star astrometric solutions and
(b) ${\tt norm\_dG}< -1$ to eliminate sources whose BP/RP colors may be corrupted by bright neighbors.

We tested these astrometric cuts on a validation set of known eclipsing WDMS binaries from \citet{Brown23}, ensuring that they remove spurious objects without excluding genuine PCEBs. Note that the additional quality filters from \citet{Rybizki22} remove only $5{,}000$ ($0.65\%$) sources but improve the purity of the CMD gap region.

With this cleaned $200$~pc {\it Gaia} sample at hand, we first correct all sources for extinction and reddening using the \citet{Edenhofer24} dust map. After these corrections, we identify unresolved WDMS candidates as those lying between two empirical CMD boundaries:
$M_G < 3.25\,(G_{\rm BP}-G_{\rm RP}) + 9.625$ and
$M_G > 4\,(G_{\rm BP}-G_{\rm RP}) + 2.1$. Not all of these stars are unresolved WDMS; in fact, a significant fraction in the upper portion of this region are expected to be low-metallicity MS stars. 

\begin{figure}
\centering
\includegraphics[width=0.99\columnwidth]{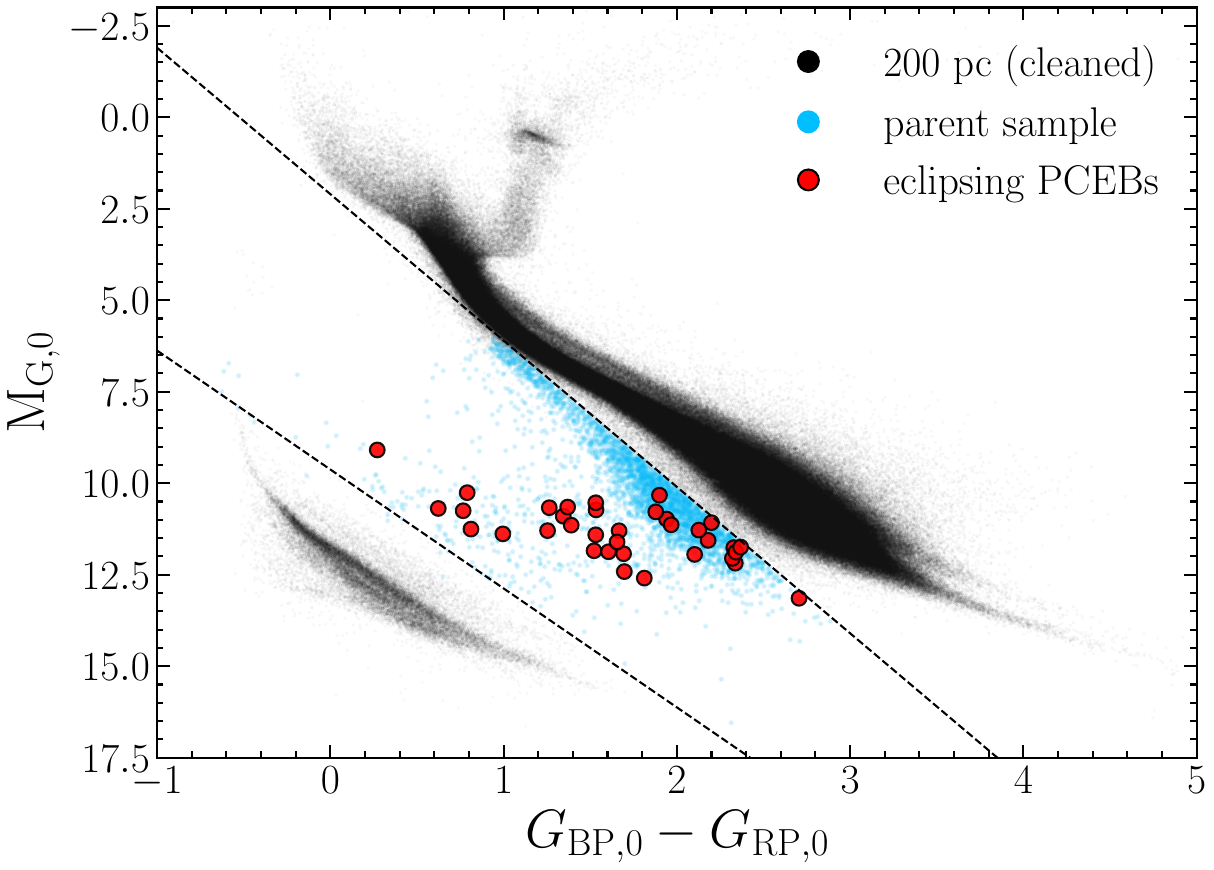}
\caption{
Sample selection in the \textit{Gaia} color--magnitude diagram.
Black points show the full $200$~pc sample after our astrometric quality cuts and extinction corrections.
The dashed lines indicate the region used to define unresolved WDMS candidates. Inside this region, we identify $N=3777$ sources (blue points) that form the parent WDMS sample, from which we identify $39$ eclipsing PCEBs (red).\label{fig:gaia_cuts}
}
\end{figure}

Figure \ref{fig:gaia_cuts} illustrates the CMD selection, showing the full \textit{Gaia} $200$~pc sample (black) and the WDMS selection region (dashed lines). The resulting candidate list of $N=3{,}777$ sources (blue) is then cross-matched with ZTF using a $2''$ positional tolerance to obtain the $g-$ and $r-$ band light curves used in our eclipse search. Among these, we identify $39$ detached eclipsing binaries (red; see Section \ref{subsec:EB_search}).

\subsection{Completeness of the parent sample}
\label{subsubsec:parent_sample_completeness}

To determine which unresolved WDMS binaries can enter our parent sample, we simulate a synthetic WDMS binary population and pass the systems through our {\it Gaia} selection criteria. This mock-observing process allows us to quantify the completeness of the parent sample with respect to the underlying WDMS population. For a detailed discussion on the types of PCEBs that our {\it Gaia} CMD search is sensitive to, please refer to Appendix \ref{app:sensitivity}. We summarize the key aspects below.

We generate an initial population of $540{,}000$ MSMS binaries using the {\tt COSMIC} binary population synthesis code \citep{COSMIC}. Initial orbital periods, primary masses, mass ratios, and eccentricities are sampled covariantly using {\tt COSMIC}'s {\tt multidim.py} module \citep{Moe17}, ensuring that the generated binary population reproduces empirical multiplicity statistics. We modify the sampler to incorporate the observed M dwarf multiplicity distribution of \citet{Winters19}, replacing the default assumption of zero multiplicity at $0.08~M_\odot$ to a $\sim20\%$ binary fraction at this mass. The modified sampling code is publicly available\footnote{\url{https://github.com/cheyanneshariat/gaia_triples}, see also \citet{Shariat25_10k}.}.

We assume a constant star-formation history by evolving each system for a randomly drawn age, chosen uniformly between $0$ and $12$~Gyr. Metallicities are sampled randomly from MS stars within $200$~pc, where [M/H] are derived from \citet{Andrae23}. We adopt the default {\tt COSMIC} mass-transfer prescriptions but set the common-envelope efficiency to $\alpha_{\rm CE}=0.3$, based on evidence for inefficient CEE in various classes of post--CEE systems \citep[e.g.,][]{Zorotovic10,Toonen13,Camacho14,Cojocaru17,Ge22,Hernandez22b,Zorotovic22,Scherbak23,Ge24,Santos25}. We also run three different {\tt COSMIC} binary populations assuming a different magnetic braking (MB) law each time: no MB (${\tt htpmb}=-1$), the \citet{Rappaport83} prescription (${\tt htpmb}=0$), and the \citet{Ivanova03} prescription (${\tt htpmb}=1$). The latter model is used for completeness estimation, but we find that this choice has only a small effect. The main differences arise in the \citet{Rappaport83} model, which predicts stronger angular–momentum losses for systems with secondary masses $\gtrsim 0.4~{\rm M_\odot}$ in this period range, causing a larger fraction of binaries to enter mass transfer.

After evolving the population, the simulation contains $\sim23{,}700$ WDMS binaries. To evaluate which systems would appear unresolved by \emph{Gaia}, we place each synthetic binary at a random distance within $200$~pc, assuming a constant space density. We approximate the projected separation by the semi-major axis $a$ \citep{Dupuy11} and require an angular separation $(a/{\rm au})(d/{\rm pc})^{-1}<2''$, roughly corresponding to \emph{Gaia}'s angular resolution for unequal-flux pairs \citep[e.g.,][]{EB_review}. 
This yields $\sim14{,}700$ unresolved WDMS binaries.

For each unresolved WDMS binary, we synthesize \emph{Gaia} $G$, $G_{\rm BP}$, and $G_{\rm RP}$ photometry by combining model MS and WD fluxes.  
For the MS component, we interpolate {\tt PARSEC v2.0} isochrones \citep{Bressan12,Chen14,Chen15,Nguyen22} and select the closest metallicity track. We then identify the nearest stellar mass and adopt its predicted \emph{Gaia} absolute magnitudes.
For the WD component, we interpolate MIST WD cooling tracks \citep{Bauer25}, using the reported WD masses and cooling ages to derive synthetic photometry. Given the WD mass and the cooling age from {\tt COSMIC}, we obtain the effective temperature, radius, luminosity, and synthetic \emph{Gaia} absolute magnitudes. 
We convert magnitudes to fluxes, sum the MS and WD fluxes in each band, and convert back to magnitudes to obtain the combined unresolved WDMS photometry. 

Finally, we require that ${\tt parallax\_over\_error}>10$ to match the {\it Gaia} selection cut (Section \ref{subsec:parent_sample}). The ${\tt parallax\_error}$ for each synthetic binary is estimated by its apparent $G$ magnitude \citep[see Table 4 of][]{Lindegren21}. At the maximum distance of $200$~pc, the ${\tt parallax\_over\_error}$ requirement corresponds to an absolute magnitude limit of $M_G\simeq13.5$ in {\it Gaia} DR3, which for an MS star implies a mass of $\approx0.12~{\rm M_\odot}$, close to the hydrogen-burning limit. $100\%$ of PCEBs and unresolved WDMS binaries satisfy ${\tt parallax\_over\_error}>10$ in the synthetic population. Thus, the ${\tt parallax\_over\_error}$ cut does not significantly bias our sample against low-mass companions. After deriving synthetic photometry, we place each system on the CMD and apply the same selection cuts used to construct our real $200$~pc {\it Gaia} sample.

Among the $14{,}700$ unresolved WDMS binaries, $2556$ ($17\%$) are detached PCEBs with $0.1 < P_{\rm orb}/{\rm day}<2$. 
Among the remaining unresolved WDMS binaries that are not PCEBs in our period range, $86\%$ were too wide to ever interact, $12\%$ are mass transfer products (stable or unstable) with $P_{\rm orb}>2$~days, and $2\%$ are presently mass-transferring CVs.

Out of all unresolved WDMS binaries, only $4\%$ fall within our CMD selection region. For detached PCEBs with M dwarf companions (defined here as $M_{\rm MS}<0.4~{\rm M_\odot}$), this fraction rises to $21\%$, consistent with previous estimates \citep[e.g.,][]{RebassaMansergas21,Santos25}. The relatively low completeness to unresolved WDMS binaries reflects the fact that most are dominated photometrically by the MS companion and lie on the MS locus rather than in the region between the WD cooling track and MS.
The completeness increases substantially for hotter WDs and lower-mass secondaries: restricting to PCEBs with $M_{\rm MS}<0.4~{\rm M_\odot}$ 
and $T_{\rm eff,WD}>9000$~K yields a completeness of $90\%$, with the rest lying blueward of our selected CMD region (i.e., the WD dominates the optical flux).
For $M_{\rm MS}<0.5~{\rm M_\odot}$ and $T_{\rm eff,WD}>9000$~K, the completeness is $70\%$, with most missed sources now being too red for the CMD selection. See Appendix \ref{app:sensitivity} for a complete outline of our search's sensitivity.

In summary, our {\it Gaia} CMD selection (Section \ref{subsec:parent_sample}) identifies only $\sim21\%$ of all unresolved WD + M dwarf PCEBs in the synthetic population, but is significantly more complete for systems with young, hot WDs. Namely, for $T_{\rm eff,WD}>9000~{\rm K}$, we estimate a completeness of $90\%$ ($70\%$) to PCEBs with $M_{\rm MS}<0.4~{\rm M_\odot}$ ($M_{\rm MS}<0.5~{\rm M_\odot}$). 
At the low-mass end, the blue CMD cut biases against very faint companions, mainly those below the hydrogen burning limit. Using the synthetic population, we find that the sample remains sensitive down to companions of $M_{\rm MS}\simeq0.1~{\rm M_\odot}$. Whether such systems enter the CMD-selected region is governed primarily by the WD temperature: very hot WDs shift systems blueward of the CMD cut, while (more commonly) cool WDs shift them redward.
Because the majority of unresolved WDMS binaries and PCEBs remain outside our CMD-selected region, our parent sample of unresolved WDMS candidates represents only a small but well-defined subset of the underlying population.

\subsection{Eclipsing binary search}\label{subsec:EB_search}

\begin{figure*}
\centering
\includegraphics[width=0.99\textwidth]{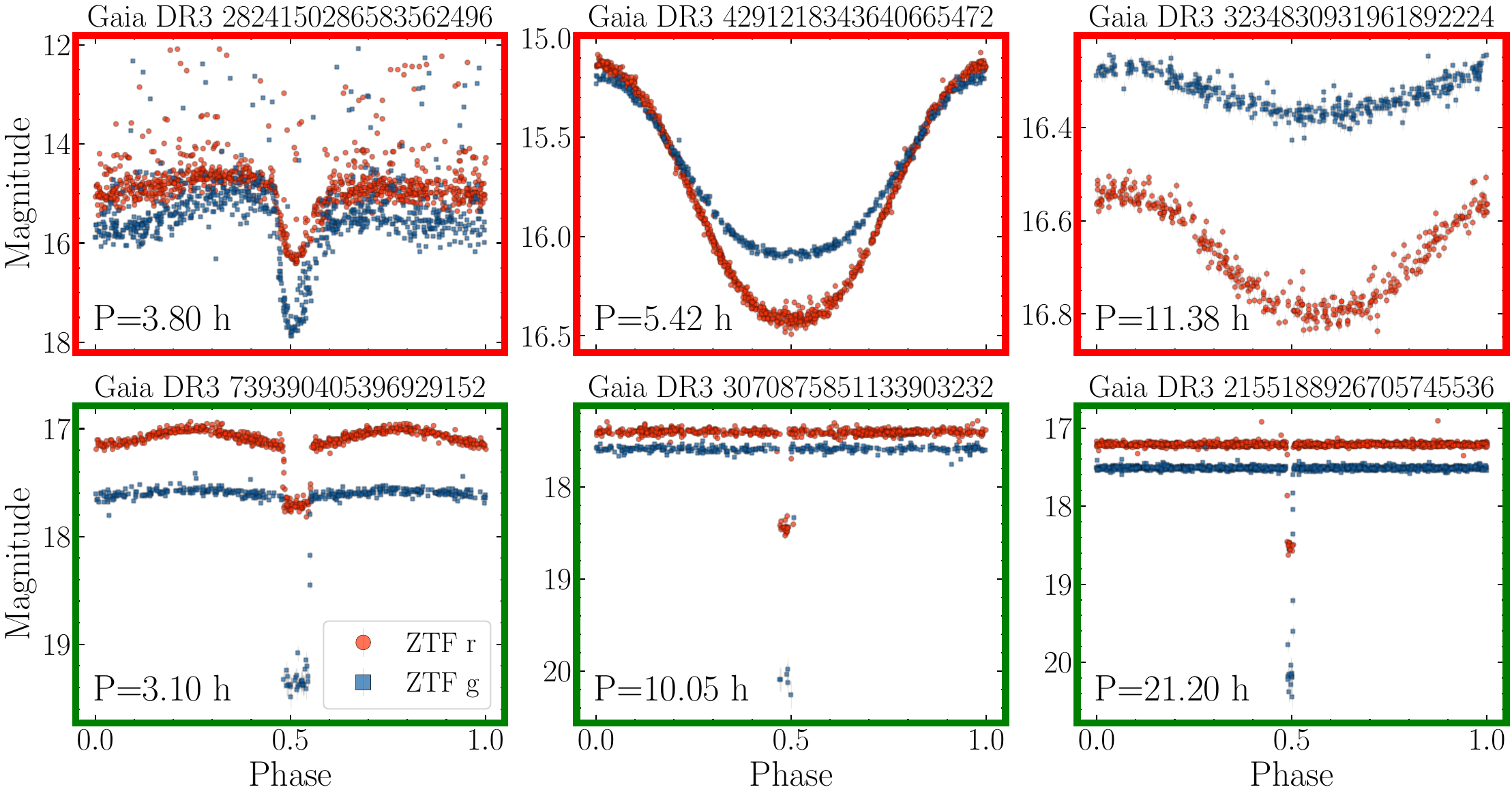}
\caption{ Periodic variables that are (bottom) and are not (top) selected in the WDMS EB sample during visual inspection. 
We show ZTF $r$-band (red circles) and $g$-band (blue squares) light curves for six representative systems. The top three are not selected, either because their variability is not due to a WD eclipse (latter two) or because they are eclipsing but are a cataclysmic variable (left, see text).
The bottom panel shows systems that are selected as WDMS EBs, which exhibit a range of eclipse depths and durations from orbital periods of $0.1$ to $2$~days. \label{fig:example_LCs}}
\end{figure*}

We searched for periodic variability in the ZTF light curves of all unresolved WDMS binaries in our parent sample ($N=3777$) using both the Fast Periodicity Weighting \citep[FPW;][]{FPW} and Box Least Squares \citep[BLS;][]{Kovacs02} periodograms. First, we normalize and combine the $g$- and $r$-band photometry into a single light curve to increase the statistical power of our periodicity search while preserving the original cadence and uncertainties. While the exact light curve shape can differ between the $g$ and $r$ bands, the underlying orbital period should be identical, so combining the bands in this way strengthens the relevant periodic signal for our purposes. After combining, we compute the FPW power spectrum for all sources over periods of $2-48$~hrs, oversampled by a factor of $30$. We avoid searching longer periods because the eclipse probability is vanishingly small for WDMS binaries, and avoid shorter periods because most systems there will be mass transferring. Known terrestrial aliases (e.g., $|(P_{\rm orb} - 1~{\rm day})/N| < 0.005$ for $N\in1,2,3,...,6$) were masked to reduce false detections. Although this may exclude a small number of genuine systems, the effect is included in our selection-function modeling and does not bias our results.

We identified the three periods with the highest FPW power for each source, folded the light curve on each period (and on its $0.5\times$ and $2\times$ aliases), and visually inspected the folded curves to look for eclipse morphology, depth, and duration consistent with WDMS binaries. Once the orbital period was identified, we refined the period by performing a BLS periodogram around the FPW period peak while exploring a range of transit durations, using the {\tt astropy} BLS implementation \citep{Kovacs02,Hartman16,astropy2018}.
The final period reported corresponds to the highest BLS likelihood.

Among the $3777$ unresolved WDMS candidates with ZTF light curves, this procedure yielded $39$ secure WDMS eclipsing binaries that are likely detached. We exclude Cataclysmic Variables (CVs) by removing systems showing outbursts or strong irregular variability. Such variability complicates completeness estimates and makes eclipse recovery more difficult. Excluding CVs also ensures the sample represents detached WDMS binaries whose orbital periods more directly trace their immediate post--common-envelope configurations. Our final sample contains $39$ detached systems with $P_{\rm orb}\in 2-36$~hr. $7$ binaries have orbital periods of $2-3$ hours and could be post--mass-transfer systems that are presently in the CV period gap. However, we find (Section \ref{subsec:magnetic_braking}) that most systems at longer periods have low-mass M dwarfs that will only overflow their Roche lobes near $P_{\rm orb}=2$~hrs, so we suspect that most systems in our sample have never interacted. The full sample, along with their light curves and basic properties, is provided in Appendix \ref{app:all_LCs} and \ref{app:SED_fitting}.

Figure~\ref{fig:example_LCs} illustrates examples of ZTF light curves that were rejected (top row) and selected (bottom row) by our visual inspection process. The top panels show systems that exhibit periodic variability but were excluded from our sample for various reasons. The first system, Gaia DR3 2824150286583562496, displays an eclipse but was rejected because it is a CV, as determined by the out-of-eclipse brightening events. The other two sources (Gaia DR3 4291218343640665472 and 3234830931961892224) show smooth, quasi-sinusoidal variations inconsistent with the sharp, short-duration eclipses expected for detached WDMS binaries; this variability may instead be due to reflection/irradiation, ellipsoidal variability, or spots.

The bottom panels of Figure \ref{fig:example_LCs} show three representative systems that are included in our final sample, with orbital periods of $\approx3$, $10$, and $20$~hours. Each displays a distinct white dwarf eclipse. Gaia DR3 739390405396929152 additionally shows clear irradiation modulation of the M dwarf companion caused by the nearby hot white dwarf. Longer-period systems ($\gtrsim10$~hr) exhibit shorter eclipse duty cycles but still display deep ($\sim1$-$3$~mag) white dwarf eclipses. Phase-folded ZTF light curves for all WDMS eclipsing binaries in our sample are provided in Appendix~\ref{app:all_LCs}.

Stellar parameters, including the effective temperature ($T_{\rm eff}$), radius ($R$), and mass ($M$) are derived for both the WD and MS components in our WDMS binary sample by performing a two-component fit to the spectral energy distribution (SED). We outline the fitting process in Appendix \ref{app:SED_fitting} and provide the resulting parameters for all systems in Table \ref{tab:WDMS_params} with a machine-readable version available online\footnote{\url{https://github.com/cheyanneshariat/pcebs}}.

\subsubsection{Are all binaries post common envelope?}
The rest of our analysis assumes these binaries are all post common envelope. 
However, hierarchical triples offer an alternative formation channel: eccentric Kozai-Lidov oscillations can drive WDMS binaries to short periods without a prior common envelope phase \citep{Kozai1962,Lidov1962,Naoz2016,Toonen2016,Toonen20,Shariat23,Knigge22,Shariat25CV}. This mechanism has been predicted to contribute $\sim20\%$ of the CV population \citep{Shariat25CV} and generally delivers systems to periods where tidal dissipation dominates (near the Roche limit), corresponding to $P_{\rm orb} \approx 1.5-3$~hr for an M dwarf companion assuming an equilibrium tidal model \citep{Hut80,Eggleton98,Kiseleva98}.
A cross-match with \citet{EB21_widebin} shows that $4/39$ ($10\%$) of our systems have resolved tertiary companions (Gaia DR3 277674209829749504, 2025873096433233664, 2562060180904793344, 433584335876697216), with an additional fraction of unresolved tertiaries expected to exist \citep{Shariat25CV}. All $4$ inner binaries in our sample have $P_{\rm orb} > 3.5$~hr, placing them above the period regime predicted by the triple channel, assuming classical equilibrium-tides \citep[though tidal heating may enlarge the circularization period;][]{Currie17,EstrellaTrujillo23}. These systems are therefore more likely to be genuine PCEB products, but the role of triples in the PCEB population deserves further study.

\subsection{Completeness of the eclipsing binary search}\label{subsec:completeness}

In Section \ref{subsubsec:parent_sample_completeness}, we quantify the completeness of our {\it Gaia} search: the types of PCEBs that satisfy our initial selection.
In this section, we characterize the completeness of our eclipsing-binary search by performing an injection-recovery experiment designed to quantify this aspect of our selection function. 

We began by generating a grid of synthetic WDMS light curves using the {\tt ellc} light curve synthesis code \citep{Maxted16}. The goal of these simulations is to obtain eclipse depths and duty cycles consistent with the range of PCEBs. Each mock system is assigned a WD and MS star with representative parameters sampled from our mock population of PCEBs (see Section \ref{subsubsec:parent_sample_completeness}), including $M_{\rm WD}$, $R_{\rm WD}$, $T_{\rm eff, WD}$, $M_{\rm MS}$, $R_{\rm MS}$, and $T_{\rm eff, MS}$. We only consider systems that fall in the CMD-selected region, and thus have some detectable flux from both the WD and MS star.
These WDs mostly exhibit $T_{\rm eff, WD} \gsim 7500$~K, with a tail out to $\sim40{,}000$~K (Appendix \ref{app:sensitivity}). The MS stars range from $0.2-1.0~{\rm M_\odot}$ with a median of $0.3~{\rm M_\odot}$ (Appendix \ref{app:sensitivity}).
Orbital periods are drawn uniformly between $0.1$ and $2.0$~days to densely sample the entire period range. Therefore, the inferred selection function does not depend on the period distribution of the mock population. Line-of-sight inclinations are sampled from an isotropic distribution ($p(i){\rm d}i = \sin i~{\rm d}i$). Each of these samples provides an {\tt ellc} theoretical light curve, of which we generate $10{,}000$.

For each simulated system satisfying the geometric eclipse condition, $\cos i < (R_1 + R_2)/a$, we inject the synthetic eclipse profiles into a control sample of ZTF light curves. To construct the control sample of ZTF light curves, we search a $5$ arcminute radius around each WDMS EB in our observed sample and selected all ZTF sources within $0.5$~mag in the {\it Gaia} $G$ band.
Repeating this for each WDMS EB candidate yielded a control set of $\approx1{,}000$ light curves. 
Using nearby stars as a control set ensures that the mock light curves share the same ZTF field and thus cadence, photometric noise, and seasonal visibility as the WDMS targets in our sample.
We use this control sample, rather than the WDMS light curves themselves, because the WDMS sources are more likely to exhibit unrelated periodic variability, which could bias the injection-recovery results.
We inject the eclipses from the {\tt ellc} model into the control stars' light curves at the ZTF cadence for the $895/10000$ of the generated mock binaries that satisfy the geometric eclipse condition.

The resulting mock WDMS light curves were passed through the same period-finding pipeline as the real sample (see Section \ref{subsec:EB_search}), including a visual inspection.
A simulated system is classified as recovered if a significant periodic signal was detected near the injected period and the phased light curve shows the expected eclipse morphology in both $g$ and $r$ bands.
After repeating this process for all injections, we recover $692/10000$ mock ZTF EBs ($7\%$). This allows us to derive the detection efficiency as a function of orbital period, as well as our sensitivity to various inclinations and WD effective temperatures.

\begin{figure}
\centering
\includegraphics[width=0.89\columnwidth]{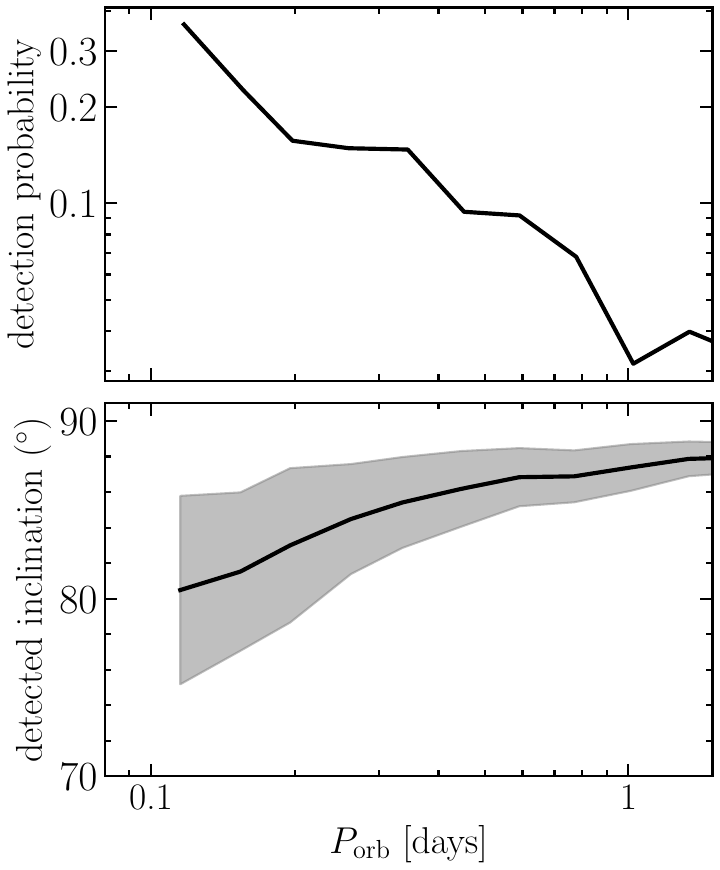}
\caption{
Sensitivity of our eclipsing binary search. We show the results of our injection recovery analysis, where the top panel shows the detection probability as a function of orbital period for all sources, and the bottom panel shows the median inclination of detected eclipsing binaries (black line with $1\sigma$ shaded). At the shortest periods, the search is sensitive to a broader range of inclinations, whereas only edge-on systems are retrieved at $P_{\rm orb} \gsim 1$~day. Even at the shortest periods, only $\sim30\%$ of systems are eclipsing. The drop at $P_{\rm orb} \approx 1$~day corresponds to the terrestrial alias.
}
\label{fig:completeness}
\end{figure}

Figure \ref{fig:completeness} summarizes the completeness of our eclipsing-binary search for systems already in the target CMD region. The top panel shows the detection probability as a function of orbital period, marginalized over eclipse depths and duty cycles. The bottom panel shows the mean line-of-sight inclination of detected systems as a function of period for EBs, with shaded $1\sigma$ intervals.

At short periods ($P_{\rm orb}\lesssim5$ hr), the search remains sensitive down to inclinations of $\sim80^\circ$, but at $P_{\rm orb}\sim1$ day, only systems that are virtually edge-on ($88^\circ \leq i \leq 90^\circ$) are detected. Even at the shortest periods, only about $30\%$ of the binaries are recovered, reflecting the low geometric probability of eclipse even for compact binaries. At fixed component radii and masses, this probability scales approximately as $P_{\rm orb}^{-2/3}$, which largely governs the observed decline in detection efficiency with increasing period.

After performing this injection-recovery exercise, we have quantified the completeness of our eclipsing binary search to PCEBs with $P_{\rm orb}\lsim2$~days.
Together with the completeness of our parent sample, our complete selection function is now well-characterized, allowing us to use our observed characteristics to explore intrinsic properties of the PCEB population.

\section{Results}\label{sec:results}
Equipped with an observed WDMS eclipsing binary sample (Section \ref{subsec:EB_search}) and a well-understood selection function (Section \ref{subsubsec:parent_sample_completeness} and \ref{subsec:completeness}), we now correct our observed sample for incompleteness and study the intrinsic PCEB population. 

\subsection{Period distribution}\label{subsec:period_dist}

\begin{figure*}
\centering
\includegraphics[width=0.7\textwidth]{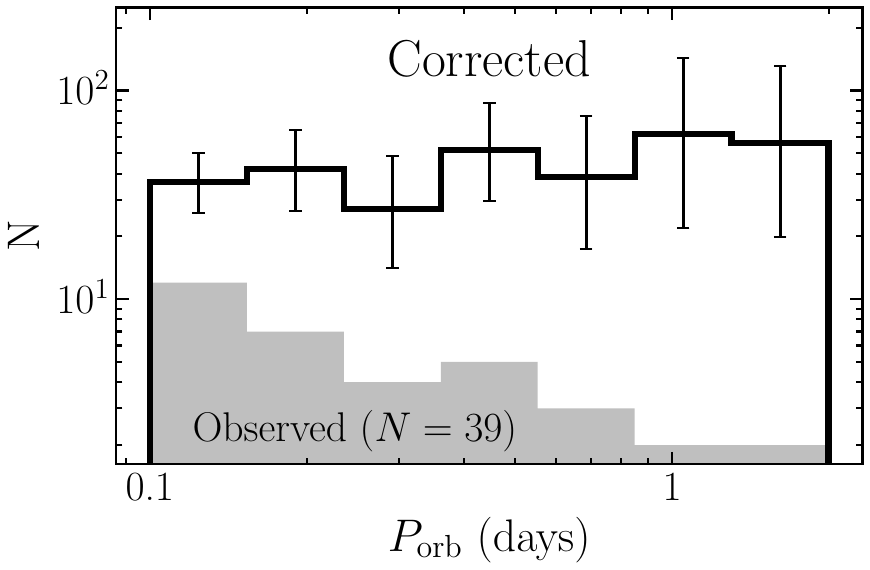}
\caption{
Completeness-corrected orbital period distribution of PCEBs. The gray histogram shows the observed period distribution of our sample. The black curve shows the bias-corrected distribution with Poisson error bars included for each bin. The orbital period distribution of WDMS PCEBs in this period range is consistent with log-uniform.
}
\label{fig:period_dist}
\end{figure*}

The intrinsic period distribution of short-period PCEBs is inferred by dividing the observed orbital period distribution by the detection probability derived in Section \ref{subsec:completeness} and shown in Figure \ref{fig:completeness}. Because the WDs necessarily contribute a significant fraction of the optical flux in the parent sample (Section \ref{subsec:parent_sample}), the detection probability is set mainly by the eclipse probability and duty cycle.

Figure \ref{fig:period_dist} shows the intrinsic period distribution of PCEBs in our sample. The gray histogram represents the observed WDMS eclipsing binaries, while the black curve shows the bias-corrected distribution with $1\sigma$ Poisson uncertainties derived using Astropy's ${\tt poisson\_conf\_interval}$ module \citep{Maxwell11}. After accounting for detection efficiency, the intrinsic distribution is consistent with being log-uniform between $0.1$ and $2$~days. 

Our observed sample only contains $4$ binaries with $P_{\rm orb}>20$~hrs. Consequently, the period distribution at $P_{\rm orb}\gsim1$~day is poorly constrained by our data, while $P_{\rm orb}\lsim1$~day is relatively well constrained and consistent with log-uniform. 

The incompleteness-correction implies that the {\it total} number of detached PCEBs in our parent sample with $0.1 \leq P_{\rm orb}/{\rm day} \leq 2$~days is $296 \pm 65$, given that only $N=39$ EBs are recovered in ZTF ($f_{\rm eclipsing} \approx 0.13$). The eclipsing fraction is consistent with the $12\%$ found by \citet{Parsons13} using a similar survey and a slightly broader period range than adopted here.

The orbital periods represented here likely reflect the immediate post--common-envelope configuration, given the sample’s bias toward young (hot) WDs. We estimate cooling ages (Appendix \ref{app:SED_fitting}) and find that most systems are within a few hundred Myr of WD formation and thus envelope ejection. Over this timescale, orbital decay from magnetic braking is minimal (Section \ref{subsubsec:period_decay_MB}). Figure \ref{fig:period_dist} therefore provides a close approximation to the intrinsic post--common-envelope period distribution of WD + M dwarf binaries.
The log-uniform period distribution inferred from our sample is notably different than the previously-reported log-normal \citep{GomezMoran11} or bimodal \citep{Shani25} distributions across a similar period and mass range (see discussion in Section \ref{subsec:comp_to_previous}).

The period distribution of PCEBs in our {\it simulated} binary population (Section \ref{subsubsec:parent_sample_completeness}) that reside in the CMD-selected region is log-uniform for models with no MB and \citet{Ivanova03} MB, but peaked for the \citet{Rappaport83} MB model. The first two models have relatively weak MB at these PCEB masses and periods, while the latter shrinks binary orbits more rapidly and preferentially at shorter periods, leading to a peak at $\approx0.4$~days. The distribution of temperature, mass, and radius for both the WD and MS companions in CMD-selected PCEBs is similar between all of these models.
For the purposes of estimating completeness, the intrinsic period distribution of a given model has virtually no impact on our estimation of the selection function. In the injection–recovery analysis (Section \ref{subsec:completeness}), orbital periods are sampled uniformly over $0.1$–$2$~days to ensure coverage over the entire range, independent of the mock population's period distribution. Then, the detection probability is determined in each bin of orbital period. 
In a given period bin, the selection function is set entirely by the effective temperature and radius of the WD and main-sequence companion, which determines whether a system enters the CMD-selected parent sample and whether its eclipse is detectable in ZTF (Figure~\ref{fig:completeness}). 
We therefore adopt the no-MB prescription for completeness estimation, but note that the resulting period distribution shape is insensitive to this choice, given the similarity in stellar properties between the models. We note that, while assuming a stronger MB law will not affect the inferred shape of the period distribution (found to be log-uniform), it may affect the normalization since stronger MB would decrease the detached lifetime of PCEBs (see discussion in Section \ref{subsec:magnetic_braking}).

\subsection{Companion mass distribution}\label{subsec:mass_dist}

\begin{figure}
\centering
\includegraphics[width=0.99\columnwidth]{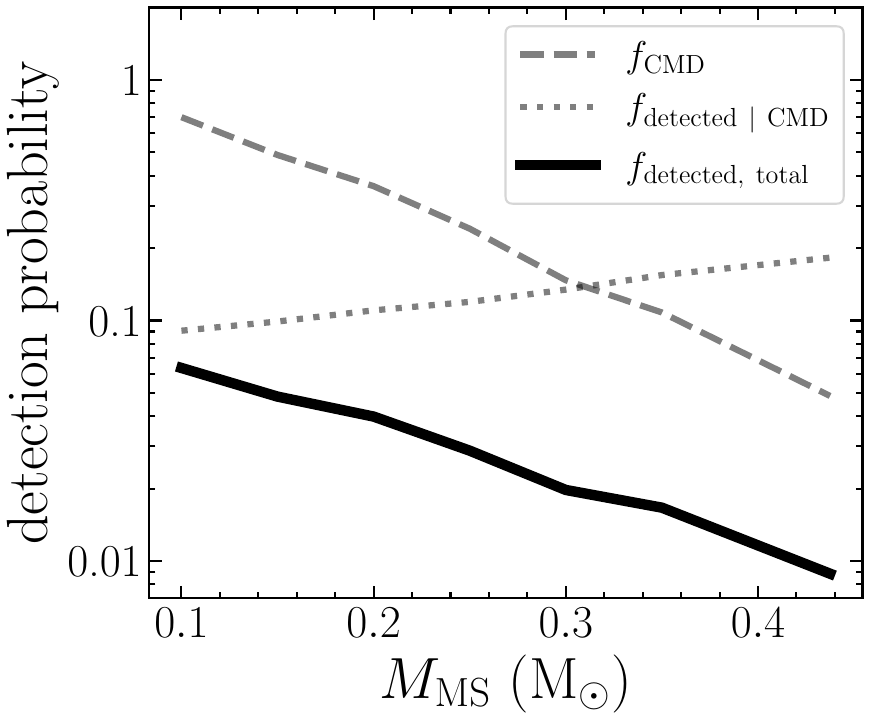}
\caption{
Detection efficiency as a function of companion mass.
The dashed curve shows the fraction of PCEBs that fall within our {\it Gaia} CMD selection region ($f_{\rm CMD}$).
The dotted curve shows the probability that a PCEB inside the CMD region is eclipsing and recovered in ZTF light curves ($f_{\rm detected~|~CMD}$).
The solid curve shows the total detection efficiency ($f_{\rm detected,~total}$).
}
\label{fig:completeness_mass}
\end{figure}

To infer the intrinsic distribution of companion masses ($M_{\rm MS}$) of short-period WD + M dwarf PCEBs, we correct the observed distribution by the mass-dependent detection efficiency of our survey. While WD masses are highly uncertain in our sample, MS masses are relatively well-constrained due to the precise parallaxes from {\it Gaia} and robust radii from SED fitting (Appendix \ref{app:SED_fitting}).

Quantifying the mass completeness depends on several coupled effects, including CMD selection, eclipse geometry, and the recoverability of eclipses in ZTF light curves.
We therefore adopt a forward-modeling approach that marginalizes over these effects using a synthetic PCEB population (see Appendix \ref{app:sensitivity}) along with our ZTF injection-recovery tests (Section \ref{subsec:completeness}).

For each $M_{\rm MS}$ bin, we estimate the total detection probability as
\begin{equation}\label{eq:mass_completeness}
    f_{\rm detected,~total} = f_{\rm in~ZTF} \times
    f_{\rm CMD} \times f_{\rm detected~|~CMD},
\end{equation}
where $f_{\rm in~ZTF}$ is the fraction of PCEBs that lie within the ZTF footprint and are unsaturated ($\delta>-28^\circ$ and $G>12.5$~mag),
$f_{\rm CMD}$ is the fraction of PCEBs that reside in the selected {\it Gaia} CMD region,
and $f_{\rm detected~|~CMD}$ is the fraction of PCEBs inside the CMD region that are both eclipsing and recovered in ZTF light curves. Note that $f_{\rm in~ZTF}$ accounts only for sky coverage and saturation limits, while all ZTF systematics, including spatially-dependent cadence and noise properties, are incorporated into $f_{\rm detected~|~CMD}$ through the injection--recovery simulations described in Section~\ref{subsec:completeness}.

$f_{\rm in~ZTF}=0.66$ because ZTF covers $\sim2/3$ of the sky and effectively all ($99.9\%$) of our WDMS parent sample are fainter than the ZTF saturation limit of $G\approx12.5$~mag.
$f_{\rm CMD}$ is evaluated directly from the synthetic population by computing, in each mass bin, the fraction of systems that fall within our CMD selection region. 
$f_{\rm detected~|~CMD}$ is derived by generating synthetic light curves for PCEBs in our population that already reside in the CMD-selected region.
We adopt a fiducial {\tt COSMIC} model with no magnetic braking below the fully convective boundary and a critical mass ratio for unstable mass transfer of $q_{\rm crit}=0.4$ \citep[e.g.,][]{Yamaguchi25}.
We experiment with alternative magnetic braking prescriptions and mass-transfer stability criteria, finding that these choices primarily affect completeness estimates for companions with $M_{\rm MS}\gtrsim0.35~{\rm M_\odot}$.
Because only one observed PCEB lies in this mass range, these assumptions mainly broaden the uncertainties there and do not materially affect our conclusions.

For each PCEB in our mock population, we assign a random orbital inclination (isotropically) to determine which systems are geometrically eclipsing. For geometrically eclipsing systems, we generate synthetic light curves with {\tt ellc} (Section~\ref{subsec:completeness}) using the component fluxes and orbital periods of each binary. From these light curves, we measure the eclipse depth and duty cycle, which are then mapped onto our empirically calibrated ZTF injection--recovery plane to assign a recovery probability as a function of eclipse morphology.
Averaging these recovery probabilities within each mass bin yields $f_{\rm detected~|~CMD}$, thereby marginalizing over orbital period, inclination, WD properties, and eclipse morphology while accounting for ZTF cadence and noise properties.

Deriving $f_{\rm detected~|~CMD}$ depends on the assumed orbital period distribution within each companion-mass bin.
In the synthetic {\tt COSMIC} population, the period distribution is approximately log-uniform across most mass bins, with the exception of the highest-mass bin ($0.35-0.5~{\rm M_\odot}$), which exhibits a modest excess near $P_{\rm orb}\simeq10$~hr due to slightly larger post--CEE separations for higher-mass companions.
However, the observed PCEB sample shows no evidence for a strong $M_{\rm MS}-P_{\rm orb}$ correlation between $M_{\rm MS}$ and $P_{\rm orb}$ beyond that expected from selection effects over the mass and period range probed here.
We also repeat the analysis assuming a strictly log-uniform period distribution in all mass bins, independent of the {\tt COSMIC} output, and find that this has a negligible effect on the inferred detection efficiencies.

Figure~\ref{fig:completeness_mass} shows the detection probability as a function of companion mass. We show the individual components, $f_{\rm CMD}$ and $f_{\rm detected~|~CMD}$, along with the total detection efficiency, $f_{\rm detected~|~total}$. The fraction of PCEBs in the CMD-selected region ($f_{\rm CMD}$) decreases as a function of $M_{\rm MS}$ because more massive (brighter) M dwarfs are more likely to dominate the optical flux contribution, causing systems to fall on the main-sequence locus of the {\it Gaia} CMD rather than within our WDMS selection region.

\begin{figure*}
\centering
\includegraphics[width=0.99\textwidth]{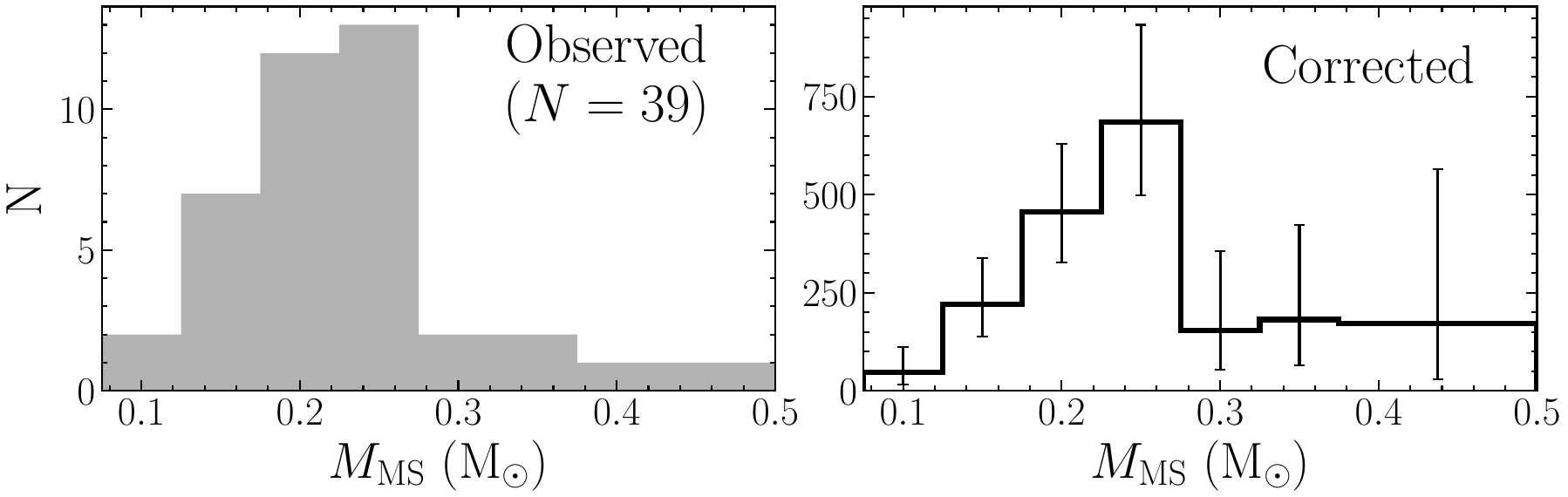}
\caption{ Completeness-corrected companion mass distribution of PCEBs within $200$~pc. {\bf Left:} Observed distribution of main-sequence star masses ($M_{\rm MS}$) in our $200$~pc sample. {\bf Right:} Corrected $M_{\rm MS}$ distribution of detached PCEBs with $P_{\rm orb}\lsim2$~days and $M_{\rm MS} \lsim 0.5~{\rm M_\odot}$ within $200$~pc. 
The intrinsic companion mass distribution of detached PCEBs peaks at $M_{\rm MS} \approx 0.25~{\rm M_\odot}$ and declines steeply in the $0.25-0.5~{\rm M_\odot}$ range.
}
\label{fig:mass_dist}
\end{figure*}

In contrast, $f_{\rm detected~|~CMD}$ increases mildly with companion mass.
This trend is driven primarily by the increasing geometric eclipse probability for larger MS stars.
For a circular orbit, the eclipse probability is $p_{\rm geom} = (R_{\rm WD} + R_{\rm MS})/a$, where $a$ is the orbital semi-major axis.
Because $R_{\rm MS} \gg R_{\rm WD}$ and assuming $R_{\rm MS} \propto M_{\rm MS}$, the eclipse probability approximately scales as $p_{\rm geom} \propto M_{\rm MS}^{2/3}$ at fixed orbital period.
Although brighter MS companions also produce shallower eclipses at fixed WD flux, this effect is mitigated by the fact that systems already reside within the CMD-selected region, so the WD contributes to the optical flux.
The combined effect of a declining $f_{\rm CMD}$ and a more weakly increasing $f_{\rm detected~|~CMD}$ results in an overall decrease in detection efficiency toward higher companion masses: from $7\%$ at $0.1~{\rm M_\odot}$ to $1\%$ at $0.4~{\rm M_\odot}$.

To correct the observed mass distribution, we divide the observed number of systems in each mass bin by the corresponding $f_{\rm detected,~total}(M_{\rm MS})$. Uncertainties are propagated by marginalizing over the realizations of the detection efficiency. 

The resulting intrinsic companion-mass distribution is shown in Figure~\ref{fig:mass_dist}, alongside the observed distribution.
The corrected distribution peaks at $M_{\rm MS} \simeq 0.25~{\rm M_\odot}$, and declines steeply at higher masses, despite the detection efficiency varying smoothly across this mass range.
Only $4/39$ ($10\%$) systems in our sample have $M_{\rm MS}\geq0.3~{\rm M_\odot}$ and only $1$ above $0.35~{\rm M_\odot}$. 
This pronounced deficit at higher companion masses, despite our sensitivity to PCEBs in this regime (Figure~\ref{fig:completeness_mass}), is a striking feature of the data and consistent with previous work \citep[e.g.,][]{Schreiber10}. We discuss its implications for magnetic braking and common-envelope evolution in Section~\ref{subsec:magnetic_braking}.

Summing all the bins in incompleteness-corrected $M_{\rm MS}$ distribution (right panel of Figure~\ref{fig:mass_dist}) provides an estimate for the total number of PCEBs inside $200$~pc with $0.1 \leq M_{\rm MS}/{\rm M_\odot} \leq 0.5$ and $0.1 \leq P_{\rm orb}/{\rm day} \leq 2$ of $N=1920^{+360}_{-320}$. Dividing the total number by the effective volume enclosed within $200$~pc (Equation \ref{eq:V_eff}) yields an estimate for the {\it local} space density of such PCEBs: $n=7.25^{+1.36}_{-1.25} \times 10^{-5}~{\rm pc^{-3}}$.
In Section \ref{subsec:space_density}, we estimate the space density of PCEBs in the Milky Way using a slightly different method and find a consistent result.

\subsection{Fraction of post common envelope binaries}\label{subsec:fraction_pceb}
Among the $3777$ sources within $200$~pc that reside in our parent sample and have ZTF coverage, we identified $39$ eclipsing binaries, corresponding to an observed fraction of $1\%$. Considering that only $13\%$ of PCEBs in the period range are recovered by our ZTF EB search (Section \ref{subsec:period_dist}), we infer that $7.6\%$ of our parent sample are detached PCEBs with orbital periods between $0.1$ and $2$~days.

The parent sample selection favors binaries where both stars contribute to the optical flux, biasing towards low-mass MS stars that are already more likely to be found in short-period orbits \citep[see also the discussion in][]{RebassaMansergas25}.
Moreover, our CMD-selected parent sample contains a significant fraction of low-metallicity MS stars that reside along the upper edge of the selected region (Figure \ref{fig:gaia_cuts}). 
Using {\tt PARSEC v2.0} isochrones \citep{Bressan12,Chen14,Chen15,Nguyen22}, we find that these contaminants are typically MS stars with [M/H]$<-1$~dex, which comprise $\sim0.3\%$ of stars in the solar neighborhood \citep[][]{Andrae23,EB24_BH3}. Hence, we expect that $2350$ of the $783211$ MS stars within $200$~pc that pass our initial {\it Gaia} query (Section \ref{subsec:parent_sample}) to reside in our CMD-selected region. 

Therefore, only $3777-2350=1427$ sources in the selected CMD region are unresolved WDMS binaries, suggesting that $21\pm4\%$ of unresolved WDMS binaries in the CMD-selected are detached PCEBs with orbital periods between $0.1$ and $2$~days. 
The estimated fraction is largely consistent with previous estimates.
An estimated $21-24\%$ of the SDSS unresolved WDMS binary sample are PCEBs over a slightly larger period \citep[out to $\sim4$ days;][]{Schreiber10,GomezMoran11}, but see also the discussion in Section \ref{subsec:SDSS}. 
Estimates from population modeling predict that PCEBs comprise $\sim30\%$ of the region between the WD and MS locus of the {\it Gaia} CMD \citep{Santos25}.

\subsection{Space density}\label{subsec:space_density}
Quantifying the space density of PCEBs allows us to place them in the broader Galactic context and provide an anchor for binary population models.
To calculate a local space density\footnote{The space density likely varies with Galactic position. We are reporting the space density at the Galactic midplane and Solar radius ($R\simeq8.1$ kpc).}, we first determine the volume enclosing a radius $d_{\rm max}$ from the Sun, accounting for the vertical density structure of the Galactic disk:
\begin{equation}\label{eq:V_eff}
    \tilde{V}(d_{\rm max}) = 2\pi \int_0^{d_{\rm max}} e^{-z/h_z}\,(d_{\rm max}^2 - z^2)~dz.
\end{equation}
We adopt a disk scale height of $h_z = 0.3$~kpc and integrate out to $d_{\rm max}=200$~pc for our rate estimation, leading to an effective volume of $\tilde{V}(200~{\rm pc}) \approx 2.6 \times 10^7~{\rm pc^3}$.
This expression approaches the total enclosed volume $4\pi d_{\rm max}^3/3$ for $d_{\rm max} \ll h_z$, but corrects for the declining stellar density with height above the Galactic plane at larger distances. Given that our entire sample resides within $d<0.2$~kpc, this correction is only mildly important.

Within $200$~pc, the {\it observed} space density of eclipsing WD + M dwarf PCEBs in our sample is\footnote{Using alternative distance limits of $d=[100,150,175]$~pc, we obtain $n_{\rm obs}=[1.0,1.5,1.75]\times10^{-6}~{\rm pc^{-3}}$ with $N=[12,21,29]$ systems in each sample, respectively. These estimates, and the $d=200$~pc value, are mutually consistent within uncertainties.} 
\begin{equation}
    n_{\rm obs} = \frac{N_{\rm obs}}{\tilde{V}(200~{\rm pc})} = \frac{39}{2.6 \times 10^7~{\rm pc^3}} =  1.5 \times 10^{-6}~{\rm pc^{-3}}.
    \end{equation}
The true space density of WD + M dwarf PCEBs with $0.1 \leq P_{\rm orb}/{\rm day} \leq 2$, $n$, is derived by correcting $n_{\rm obs}$ with our selection function. Namely,
\begin{equation}\label{eq:n_obs}
    n_{\rm obs} = n \times f_{\rm CMD} \times f_{\rm in~ZTF} \times f_{\rm eclipsing},
\end{equation}
where $f_{\rm CMD}$ is the fraction of PCEBs that satisfy our {\it Gaia} CMD cuts (i.e., reside between the WD and MS tracks), $f_{\rm in~ZTF}$ is the fraction of PCEBs that are observed by ZTF (satisfy $\delta>-28^\circ$ and $G>12.5$~mag), and $f_{\rm eclipsing}$ is the fraction of those detected as eclipsing with ZTF light curves. 

$f_{\rm CMD}=0.24$ of unresolved WD + M dwarf binaries reside in the CMD region of the parent sample (Section \ref{subsubsec:parent_sample_completeness}). $f_{\rm in~ZTF}=0.66$ because ZTF covers $\sim2/3$ of the sky and virtually all WDMS binaries in our parent sample are fainter than the ZTF saturation limit. $f_{\rm eclipsing}=0.13$ based on our injection-recovery tests (Section~\ref{subsec:completeness} and \ref{subsec:period_dist}), while the rest are either not eclipsing or are eclipsing but not recovered in the ZTF light curves. Importantly, the initial quality cuts are predicted to remove almost no genuine systems within $200$~pc (Section \ref{subsubsec:parent_sample_completeness}), so they are not included in this selection function calculation.
Together, these factors imply that only about $1$ in $60$ WD + M dwarf binaries with $0.1 < P_{\rm orb}/{\rm day} < 2$ are detected as EBs in our survey. 
Correcting for incompleteness yields an intrinsic local space density of $n = 7.2\times10^{-5}~{\rm pc^{-3}}$.

Previous estimates for post--common-envelope WDMS binaries nearly all predict densities of $n\approx 0.5-5\times 10^{-5}~{\rm pc^{-3}}$ \citep{Schreiber03, RebassaMansergas21,Inight21}, similar to our inferred density. 
A seminal study by \citet{Schreiber03} inferred the PCEB space density using the sample of PCEBs within 100 pc known at the time. To correct for incompleteness, they rescaled their observed space density by $0.5/\overline{f}_{{\rm PCEB}}$ (their Equation 12), where $\overline{f}_{{\rm PCEB}}$ is the mean fraction of the predicted total PCEB lifetime (i.e., from common envelope until future Roche lobe overflow) that has elapsed for PCEBs in their sample. This estimate assumes the PCEB population is approximately in steady state, such that the average PCEB should have $f_{\rm PCEB} = 0.5$. They estimated a space density of $2.9\times10^{-5}~{\rm pc^{-3}}$ (assuming saturated MB) or $6.6\times10^{-6}~{\rm pc^{-3}}$ (assuming RVJ MB), with the factor of $\sim4$ difference arising from the longer detached lifetimes predicted under saturated MB.

Many PCEBs in the \citet{Schreiber03} sample have predicted detached lifetimes that significantly exceed the age of the Universe. This renders the steady state assumption inappropriate, and in isolation, this assumption would lead to an {\it overestimated} PCEB space density. However, \citet{Schreiber03} also implicitly treat the {\it geometric} and {\it survey} selection of the then-known PCEB sample as approximately complete within 100 pc, applying an incompleteness correction primarily motivated by the age distribution (i.e., the lack of old/cool systems). In fact, incompleteness in their heterogeneous sample of PCEBs likely also arose from reasons unrelated to evolutionary age: the parent searches had non-uniform sky coverage, depth, and target selection criteria. This incompleteness is evidenced by the fact that only $1$ of the $12$ eclipsing PCEBs in our sample within $100$~pc -- all of which contain relatively young and bright WDs and low-mass MS stars -- were included in their sample. 

High-quality astrometry from {\it Gaia} and light curves from ZTF allow us to assemble a PCEB sample with better-understood completeness than the \citet{Schreiber03} sample. We correct our observed space density by the {\it observable} lifetime of PCEBs rather than the detached lifetime, as quantified by our selection function. Our constraints are only weakly sensitive to the assumed MB law, since we find (Section \ref{subsubsec:period_decay_MB}) that most PCEBs will not begin mass transfer in a Hubble time. Our inferred space density is a factor of $\sim 2-3$ higher than the larger of the two estimates from \citet{Schreiber03}.

Our predicted space density, $n = 7.2\times10^{-5}~{\rm pc^{-3}}$, is $20$ times higher than the space density of CVs \citep[$n \simeq 3-7\times10^{-6}~{\rm pc^{-3}}$;][]{Pala20,Inight23,Rodriguez25}, and roughly $150$ times higher than that of AM~CVn systems \citep[$\sim5\times10^{-7}~{\rm pc^{-3}}$;][]{Carter13,Rodriguez25}. The density is $\sim60$ times lower than the total space density of all WDs \citep[$4.5\times10^{-3}~{\rm pc^{-3}}$;][]{Hollands18} and $650$ times lower than the total space density of all M dwarfs assuming that half of all stars ($n_\star=0.1~{\rm pc^{-3}}$) are M dwarfs. This implies that $\approx1.7\%$ of WDs and $\approx0.15\%$ of M dwarfs reside in short-period ($0.1 \leq P_{\rm orb}/{\rm day} \leq 2$) detached PCEBs.

\subsection{Birth rate}\label{subsec:birth_rate}
The birth rate of PCEBs in the Galaxy is crucial for studying their formation efficiency, interpreting the space density, and 
placing them in the context of the broader binary population. The formation rate depends critically on the assumed observable lifetime of such systems, which we estimate using our sample's well-understood completeness.

We first estimate $N_{\rm tot}$, the total number of WD + M dwarf PCEBs with $0.1 \leq P_{\rm orb}/{\rm day} \leq 2$ in the Milky Way, from our inferred local space density. The Galactic stellar distribution is modeled as an exponential disk with vertical scale height $h_z = 0.3$~kpc and radial scale length $h_R = 3$~kpc. 
The total number of systems is then
\begin{equation}
  \begin{split}
N_{\rm tot} &= n \int_0^{2\pi} \int_{-\infty}^{\infty}  \int_0^{\infty} 
\exp \!\left(-\frac{R - R_\odot}{h_R} - \frac{|z|}{h_z}\right) R\,dR\,dz\,d\phi \\
&= 4\pi n h_z h_R^2 \exp(R_\odot/h_R)  \\
&\approx 488~\left(\frac{n}{{\rm kpc^3}}\right),
  \end{split}\label{eq:Ntot}
\end{equation}
where $n = 7.2\times10^4~{\rm kpc^{-3}}$ is the local density of WDMS PCEBs, and $R_\odot = 8$~kpc is the Sun’s Galactocentric distance. Hence, the total estimated number of detached PCEBs in the Galaxy with $0.1 \leq P_{\rm orb}/{\rm day} \leq 2$ is $N_{\rm tot}\approx3.5\times10^7$.

From our synthetic PCEB population (Section \ref{subsubsec:parent_sample_completeness}), we can determine the lifetimes of present-day PCEBs ($\tau$), defined as the time elapsed since common envelope. The lifetime distribution is broad, with $\langle \tau \rangle = 4.25$~Gyr and $\sigma_{\tau} = 2.8$~Gyr, so we estimate the birth rate using the median lifetime and uncertainties with the 16th and 84th percentile, finding
\begin{equation}
    \mathcal{R}_{\rm MW} = \frac{N_{\rm tot}}{\tau} 
    \approx 1.25
    ^{+3.65}_{-0.63}\times10^{-2}~{\rm yr^{-1}}.
    \end{equation}
This birth rate implies that a new WD + M dwarf binary emerges from common envelope at these periods roughly every $80$ years in the Milky Way. 

\subsubsection{Comparison to LRNe and ELM WDs}
The predicted birth rate of PCEBs from our sample is $\sim10\times$ lower than the total observed rate of luminous red novae (LRNe) in the Milky Way \citep[$\sim0.1-0.5~{\rm yr^{-1}}$;][]{Kochanek14,Howitt20}. The difference is sensible given that our observed sample (a) only contains a subset of detached PCEBs with a specific range of companion masses ($\lsim0.4~{\rm M_\odot}$) and orbital periods ($\lsim2$~days) and (b) only accounts for PCEBs that survive common envelope (i.e., do not merge), whereas LRNe likely include failed common envelopes (i.e., stellar mergers) as well \citep[e.g.,][]{Kochanek14,Pejcha14}.

Observations of extremely low-mass (ELM) WDs  ($M_{\rm WD} \lesssim 0.3~{\rm M_\odot}$) may provide useful context for understanding the outcomes of CEE with mass ratios similar to those studied in this work. ELM WDs ($M_{\rm WD} \lesssim 0.3~{\rm M_\odot}$) are understood to form exclusively through binary evolution, since single-star evolution cannot produce such low-mass WDs within the age of the Universe. While some ELM WDs are likely produced by stable mass transfer, their contribution to the total birth rate is small \citep[e.g.,][]{Brown16,EB21_ELM}, and the majority of ELM WDs are born after CEE.

A key result of the ELM survey \citep{Brown10, Brown20} is that, after correcting for observational biases, most ELM WDs emerge from CEE at periods of $P_{\rm orb} \lesssim 1$~hr \citep{Brown16}. 
At such separations, a main-sequence star of comparable mass would overfill its Roche lobe and therefore be expected to merge during or shortly after CEE.
Because the WD masses in ELM binaries are comparable to those in our WD+M dwarf PCEB sample, this comparison might suggest that a significant fraction of CEE events at these mass ratios lead to mergers rather than surviving  PCEBs.  

However, the comparison is not one-to-one, given that the CEE episode in ELM formation typically involves a WD inspiraling through the envelope of a giant with a low-mass core (the ELM WD progenitor), which may imply systematically smaller pre-CEE separations and possibly more tightly bound envelopes than in WD+M-dwarf progenitors. A more direct comparison would require observational constraints on binaries containing ELM WDs with M dwarf companions, which remains an important avenue for future work.

\section{Discussion}\label{sec:discussion}

\subsection{Magnetic braking}\label{subsec:magnetic_braking}

For detached WD + M dwarf binaries with orbital periods in the range spanned by our PCEB sample ($0.1-2$~days), the primary mechanism driving orbital decay is magnetic braking (MB). In the following sections, we examine the amount of MB-driven orbital decay experienced by the binaries in our sample since the common envelope. Then, we discuss using samples of detached PCEBs, like the one constructed here, to empirically constrain MB models. Details of MB modeling are provided in Appendix \ref{app:MB_equations}.

\subsubsection{Period decay since common envelope}\label{subsubsec:period_decay_MB}

Most white dwarfs in our sample have effective temperatures of $8{,}000-20{,}000$~K, corresponding to cooling ages (or equivalently the time since CEE ended) of $\sim60$-$1000$~Myr \citep{Bauer25}. For low-mass M dwarfs, which comprise the companions of our PCEB sample, observations show that magnetic activity and angular momentum loss saturate above a critical rotation rate \citep[e.g.,][]{Stauffer94,Delfosse98,Reiners09,Newton17,Johnstone21,ElBadry22_satMB, Belloni24a}. In particular, the observed period distribution of close MSMS binaries ($P_{\rm orb}<10$~days) at low masses ($0.1\lsim M/{\rm M_\odot}\lsim0.9$) is consistent with an MB law that saturates at short periods \citep{ElBadry22_satMB}.
Motivated by observational support for saturated MB from both singles and binaries, we first consider a saturated MB prescription of \citet{Sills00}.

Using the \citet{Sills00} saturated MB law (hereafter S00), we find that orbital decay over the typical WD cooling ages in our sample is modest (see Appendix \ref{app:MB_equations} for full MB equations). The median cooling age for PCEBs with UV photometry, where WD effective temperatures are more reliable, is $600$~Myr. Assuming an initial period of $12$~hrs and typical WD (MS) masses of $0.5~{\rm M_\odot}$ ($0.3~{\rm M_\odot}$), this corresponds to only an $8\%$ change in the orbital period due to MB.
Even for the oldest systems ($t_{\rm cool}\approx1.5$~Gyr), the fractional period decrease is $19\%$. The $\Delta P_{\rm orb}$ due to MB is smaller than, or at most of order, the bin size of our period distribution (Figure \ref{fig:period_dist}).

We also examine the classical RVJ MB prescription \citep{Verbunt81,Rappaport83}, derived from G-type stars with rotation periods of $2-30$~days. In most close binaries where MB significantly drives the orbital evolution (e.g., CVs and LMXBs), the donor is of spectral type K-M. Therefore, applying the RVJ law for these binary classes involves substantial extrapolation beyond its original scope. Nevertheless, it has been widely adopted in previous works studying the evolution of accreting binaries and PCEBs, so we test it here as well. Assuming the RVJ law (without disruption of MB below the fully-convective boundary) predicts a period change to $8-14\%$ over the same parameter range, weaker than in the saturated case at these masses and orbital periods. \citet{Schreiber03} also conclude that most detached PCEBs in their sample are minimally affected by MB, assuming both a saturated and classical MB prescription.

For either MB model, the {\it maximum} period evolution from the above exercise is $\sim20\%$, similar to the bin width in our observed period distribution. The observed distribution in Figure \ref{fig:period_dist} is thus a relatively close reflection of the post--common-envelope orbital separations with minimal contribution from subsequent MB-driven evolution.

\subsubsection{How long before the onset of mass transfer?}\label{subsubsec:future_evolution}

To estimate which of the detached PCEBs in our sample will eventually evolve into mass–transferring CVs, we model their future orbital evolution under the RVJ \citep{Verbunt81,Rappaport83} and saturated S00 \citep{Sills00} MB formulations. For each system we compute the present–day angular momentum loss rate using the stellar masses, radii, and orbital periods (Appendix \ref{app:SED_fitting}), and integrate the corresponding $\dot{P}_{\rm orb}$ equation forward in time until the companion fills its Roche lobe using the analytic expressions in Appendix \ref{app:MB_equations}.

For the low–mass companions typical of our sample ($M_{\rm MS} \approx 0.2~{\rm M_\odot}$), the critical rotation period for MB saturation is $\approx16.8$~days \citep{Wright11}. Most observed PCEBs are below this threshold and therefore reside in the saturated regime.
Among the $39$ detached PCEBs in our observed sample, both the S00 and RVJ prescriptions predict that $32$ systems will reach Roche–lobe overflow within a Hubble time. Both prescriptions agree that every system with a present–day orbital period $\lesssim12$~hr will evolve into a CV. The predicted time until mass transfer, however, differs between the models: the RVJ law yields a median onset time of $\langle t_{\rm onset}\rangle \approx 1$~Gyr, whereas the saturated S00 law gives longer delays, with $\langle t_{\rm onset}\rangle \approx 2.5$~Gyr for our entire observed PCEB sample.

Assuming that detached PCEBs are distributed log-uniformly in $P_{\rm orb}$ between $0.1$ and $2$~days (Figure \ref{fig:period_dist}), approximately $50\%$ of systems lie below $P_{\rm orb}=12$~hr. If either the RVJ or S00 law operates, this implies that roughly half of all newly formed detached PCEBs will ultimately become CVs. Assuming a detached-PCEB birth rate of $\mathcal{R}_{\rm PCEB} = 0.01~{\rm yr^{-1}}$ (Section \ref{subsec:birth_rate}) corresponds to a predicted CV birth rate of $\mathcal{R}_{\rm CV} \approx 5\times10^{-3}~{\rm yr^{-1}}$. 
This estimate is $10\times$ larger than expected for CVs,  given the their typical space density \citep[$\approx5\times10^{-6}~{\rm pc^{-3}}$;][]{Pala20,Inight21,Inight23,Rodriguez25} and lifetime \citep[$\sim5$~Gyr; e.g.,][]{Warner95}, derived following Section \ref{subsec:birth_rate}.

This tension may be reconciled with a disrupted MB law, whereby MB shuts off or is significantly weakened below the fully convective boundary $M_{\rm MS} \leq 0.35~{\rm M_\odot}$, as proposed in several evolutionary models \citep[e.g.,][]{Rappaport83} and discussed further in Section \ref{subsubsec:constraining_MB}. Virtually all PCEBs in our sample fall in this low-mass regime (except QS Vir), where magnetic braking is expected to be suppressed or absent. In this case, orbital angular-momentum loss is dominated by gravitational-wave radiation, which is comparatively much weaker at the observed periods ($P_{\rm orb}\gtrsim2–12$ hr; \citealt{Peters64}). As a result, many systems could remain detached for a Hubble time, even if their present-day orbital periods are shorter than $\sim12$ hr.

\subsubsection{Constraining MB models}\label{subsubsec:constraining_MB}

Well-characterized samples of detached PCEBs provide valuable empirical constraints on MB prescriptions \citep[e.g.,][]{Schreiber10,RM13_MB,Belloni24a,Blomberg24}.
The companion-mass distribution is particularly diagnostic because different MB models predict different efficiencies for removing orbital angular momentum above and below the fully convective boundary \cite[e.g.,][]{Politano06}.

\citet{Schreiber10} showed that the SDSS PCEB sample is best reproduced by a disrupted MB model in which MB efficiency drops sharply below the fully convective boundary. Their observed companion-mass distribution peaks near $M_{\rm MS}\sim0.1-0.2~{\rm M_\odot}$ and declines steeply above these masses.
A similar behavior is seen in the sdB+MS PCEB population studied by \citet{Blomberg24}, whose hotter (younger) sdB primaries allow even cleaner constraints. They also find a pronounced deficit of companions above $\sim0.2~{\rm M_\odot}$, again favoring disrupted MB with a strong boost in efficiency at higher masses.
Using population-synthesis modeling, \citet{Belloni24a} found that the SDSS PCEB sample requires a saturated MB law that is disrupted and strongly boosted (by a factor $\gtrsim50$) when the secondary is above the fully convective, naturally producing a downturn in the detached PCEB fraction across this mass boundary. A follow-up study by \citet{Barraza25} suggests that this MB model is consistent with observed properties of accreting CVs as well.

Our eclipsing PCEB sample shows a relatively sharp decrease in binaries beyond $M_{\rm MS}\sim0.25~{\rm M_\odot}$ (Figure~\ref{fig:mass_dist}), consistent with previous findings \citep[e.g.,][]{Schreiber10, Blomberg24}. Only one PCEB in our sample, QS Vir, has a secondary mass above the fully convective boundary\footnote{The spectroscopic mass of QS Vir \citep[$0.382 \pm 0.006~{\rm M_\odot}$][]{Parsons16} is lower than our photometric estimate ($0.45~{\rm M_\odot}$), but still confidently above the fully convective boundary.} and only $4$ systems in total have $M_{\rm MS} \geq 0.3~{\rm M_\odot}$.
Our sensitivity simulations show that the ZTF + {\it Gaia} search is sensitive to companion masses between $0.3-0.5~{\rm M_\odot}$ (Figure \ref{fig:completeness_mass}), suggesting that the deficit of such companions in our observations is a real phenomenon. While this trend is consistent with an MB model that is disrupted below the fully convective boundary, it is unclear from our data alone whether this trend is mostly a consequence of MB or CEE.

\subsection{Comparison to previous work}\label{subsec:comp_to_previous}

\subsubsection{TESS}\label{subsec:TESS}

Recently, \citet{Shani25} searched for WDMS EBs in TESS starting from the parent catalog of \citet{Li25}, which used a neural network classifier to identify unresolved WDMS binaries from {\it Gaia} XP spectra. This paper was in part motivated by their work, but we use a different parent sample and ZTF light curves rather than TESS. Because PCEBs are relatively faint (the median G-band magnitude for our final sample is $16.8$), ZTF's higher sensitivity and angular resolution make it easier to accurately identify eclipsing PCEBs.

\citet{Shani25} report $107$ periodic variables and classified $74$ of these as WDMS ``Eclipsing" binaries. They model the selection biases of their ``eclipsing'' sample to infer properties of the PCEB population, finding a bimodal distribution of $\log(P_{\rm orb})$, with one mode peaking at $4.1$~hrs and the other rising toward long periods and peaking beyond the maximum period of $\sim2$~days probed by their sample. They also infer that $29.8\% \pm 4.5\%$ of the \citet{Li25} sample are PCEBs.

Of the $39$ EBs in our ZTF sample, $5$ also appear in the \citet{Shani25} TESS sample, with one other classified as CV (Gaia DR3 3612227169936143360). \citet{Shani25} have $7$ eclipsing detached PCEBs in their sample (see below) located within the ZTF footprint and within $200$~pc. The two that are not retrieved in our search are Gaia DR3 2919098716285083776 and Gaia DR3 1600249121650105472. The first source has sparse ZTF cadence and a shallow eclipse; although ZTF recovers the correct period ($15.9$~hr), the folded light curve does not exhibit a recognizable WDMS eclipse. Gaia DR3 1600249121650105472 also has too few points in the ZTF light curve for eclipse recognition.

Visual inspection of the TESS light curves from the \citet{Shani25} EB sample reveals significant contamination from sources that are not eclipsing WD + MS binaries. In particular,

{\it 1.} A substantial fraction ($\sim50-60\%$) of the $52$ ``eclipsing'' binaries used for their results do {\it not} exhibit WDMS eclipses, but show other forms of variability (e.g., reflection, ellipsoidal variability, rotational variability due to spots, eclipses of non-WDMS binaries, etc.)\footnote{ \fontsize{7.5}{10}\selectfont These are Gaia DR3 1780220380340013568, 2931274020848547200, 4298332012051652352, 5337884487666708096, 5375241185442342144, 5439392500602380544, 5483999652978511232, 5516175520745007744, 6130145940426733824, 1006621281985546240, 1685720070351584000, 2250058535161610240, 2287248210301342336, 2451154710754640640, 2528989654282071936, 2856883739877737216, 2937111534244417664, 3212985152042613120, 4588070505828585984, 4658696016034748928, 4661744416353375488, 4701214616008598144, 4858226697521522432, 4896134903510805888, 518137368471752960, 5216574579445522304, 5237255534212898816, 5362212384974013184, 5433756094759976192, 6509224907926516480.}.  

{\it 2.} Their classification procedure assumes that any source already listed in SIMBAD as a CV secures its EB status.  
This assumption is problematic because most CVs are not eclipsing.

{\it 3.} Their methodology allows a {\it ``pronounced, on-target secondary eclipse"} to secure EB status after attempting to model flux dilution.  
This choice leads to contamination by MSMS binaries, particularly at long periods. Because WDs are much smaller than M dwarfs, genuine WDMS binaries are essentially never expected to produce secondary eclipses detectable in TESS. While one might speculate that a secondary eclipse could result from an eclipse of the M dwarf by a disk around the WD, such disks are not expected at the long periods where they find secondary eclipses. We conjecture that the sources \citet{Shani25} find with secondary eclipses (e.g., Gaia DR3 2931274020848547200, 4661744416353375488) are misclassified MSMS binaries.

Although \citet{Shani25} carried out an extensive injection-recovery analysis to model their eclipse-detection efficiency, the procedure is difficult to reconcile with the properties of their observed sample. Their simulations injected physically realistic WD + MS eclipses, computed using {\tt PHOEBE} \citep{Prsa16}, into real TESS light curves and measured the recovery rate as a function of orbital period. They specifically sample the orbital inclination isotropically to factor in the geometric eclipse selection bias, and state that the detection curve provides the {\it ``combined probability that a WDMS binary will eclipse and be successfully identified by our detection pipeline"} \citep{Shani25}.
However, given that the majority of their observed sample does not exhibit WDMS eclipses, the derived detection curve is likely inaccurate. 
These challenges highlight the advantages of using a clean, geometrically selected eclipsing sample for population demographics of short-period PCEBs.

\subsubsection{SDSS}\label{subsec:SDSS}
\citet{GomezMoran11} used SDSS spectroscopy to identify WDMS PCEBs via radial-velocity (RV) variability. Among $1246$ WDMS binaries with at least $2$ RV measurements, they identified $191$ PCEB candidates --  sources with RV variability detected with low-resolution spectra -- and derived orbital periods for a subset through photometric and spectroscopic follow-up. Among the $13$ SDSS PCEBs within 200 pc, $8$ reside in our CMD sample, and the other $5$ lie just outside of our CMD cuts: $3$ are too red and $2$ are too blue. Among the $8$, one is eclipsing (Gaia DR3 1245093225062192640), which we also recover. Similarly, we compare to the $30$ PCEBs from \citet{Schreiber03}, of which $10$ are within $200$~pc and in the ZTF footprint. We recover $8$, where the other $2$ lie just outside our CMD cut (one redward, one blueward). Among the $8$, one is eclipsing (QS Vir) and identified in our sample (Gaia DR3 3612227169936143360).

Based on a ``core" sample of $65$ SDSS PCEBs, \citet{GomezMoran11} modeled their detection efficiency as a function of orbital period, inferring an intrinsic log-normal period distribution that peaks at $\sim10.3$~hr with a dispersion of $0.44$~dex. Notably, this distribution predicts virtually no PCEBs with $P_{\rm orb} \gsim5$~days. 
Later on, \citet{Zorotovic11b} found a statistically significant mass dependence of this period distribution, whereby lower mass WDs (mostly He WDs) had systematically shorter orbital periods than those with more massive WDs.

In contrast to the log-normal period distribution inferred by the SDSS sample, our results favor a roughly log-uniform intrinsic period distribution across $0.1-2$~days. 
We conjecture that the drop-off in the period distribution inferred by \citet{GomezMoran11} stems at least in part from incompleteness in the SDSS selection, which relied on RV variability in $2$-$3$ epochs detectable with low-resolution spectroscopy \citep{RebassaMansergas07,GomezMoran11}, making the search insensitive to longer-period systems with small velocity amplitudes. 

Longer-period PCEBs are known from other searches. By far the nearest post-interaction WDMS binary is G~203-47 \citep{Delfosse99}, which contains a $\approx 0.3~{\rm M_{\odot}}$ M dwarf in a $14.9$~day orbit\footnote{\citet{Delfosse99} report an orbital period of $14.7136 \pm 0.0005$ d. Our follow-up spectroscopy reveals a period of $P_{\rm orb}=14.909777 \pm 0.00001$~d, which is qualitatively similar to, but highly inconsistent with, this result. The reason for the erroneous period reported by \citet{Delfosse99} is unclear.} with a WD companion. According to the lognormal fit of \citet{GomezMoran11}, only 1 in $10^4$ PCEBs should have an orbital period as long as G~203-47. Yet at a distance of only $7.5$~pc, the system is one of the $10$ nearest known WDs and almost certainly the nearest PCEB. It is $3$ times nearer than the nearest short-period PCEB, RR Cae ($P_{\rm orb} = 7.2$~hr; $d=21$~pc). While population-level inference based on one object is uncertain, the existence of G~203-47 -- and other wide PCEBs at somewhat larger distances  \citep[e.g.][]{Wonnacott93, Yamaguchi24, Garbutt24} -- implies that wider PCEBs not only exist, but are likely common.

\citet{Schreiber10} used the SDSS PCEB sample to measure the fraction of PCEBs among WDMS binaries as a function of companion mass, correcting for observational biases related to spatial resolution and RV detectability. 
They found a pronounced decline in the number of PCEBs with companion masses in the range $M_{\rm MS}\simeq0.25$–$0.4~{\rm M_\odot}$, which they interpret as support for disrupted MB at the fully convective boundary. Our results are qualitatively consistent with this finding: after correcting for incompleteness, we likewise infer a steep drop in the intrinsic companion-mass distribution above $M_{\rm MS}\simeq0.25~{\rm M_\odot}$. From our sample alone, it is unclear whether this decline at higher masses is due to MB or because such companions are more likely to experience CEE.

An important limitation at the time was the lack of precise distances, which made it difficult to infer reliable companion masses except in systems with high-quality spectra. The typical distance uncertainty in the SDSS PCEB sample is $\sim50$~pc, compared to a median uncertainty of $\sim1.5$~pc for PCEBs in our {\it Gaia}-selected sample.
With accurate parallaxes, we are able to infer precise companion masses from their radii for nearly all systems in our sample.

\subsubsection{ZTF}\label{subsec:ZTF}
Several previous ZTF-based searches identified WDMS PCEBs, but with selection functions that differ from the approach taken in this work.
\citet{Keller22} cross-matched {\it Gaia} white dwarf candidates with ZTF DR3 light curves and performed a BLS search for eclipsing binaries. They discovered $18$ new WD binaries, $17$ of which are eclipsing with periods spanning up to periods up to $0.1$~day, which is our minimum period searched. While their study does quantify the detection efficiency for each source, it does not attempt to completeness-correct their entire sample.

A complementary analysis by \citet{Li24} examined ZTF light curves for $\sim200$ WDMS binaries exhibiting Balmer emission in SDSS or LAMOST spectra. They identified $55$ PCEBs with periods from $2.2$ to $81.6$ h, with $3$ eclipsing. Because they select through chromospheric emission rather than eclipses, it recovers intermediate-period systems ($\sim0.1-3$ d), but this complicates the selection function. As a result, while \citet{Li24} report valuable new discoveries, the selection biases preclude any population-level inference.

\citet{Brown23} carried out follow-up high-precision photometry of $43$ WDMS eclipsing binaries discovered by ZTF. By fitting models to the precise multi-band light curves, they derive stellar parameters for $30$ eclipsing PCEBs. The $13$ without robust parameters either contained magnetic WDs (9) or had substellar secondaries (4). \citet{Torres25} applied inverse population synthesis to this sample of $30$ to infer the progenitor properties of each binary and to constrain common-envelope processes. Their analysis favors an average CE efficiency of $\alpha_{\rm CE}\approx0.6$, but importantly finds evidence against a universal $\alpha_{\rm CE}$. Instead, the inferred efficiencies vary substantially from system to system, indicating that CE ejection may depend sensitively on the masses, evolutionary state, or internal structure of the progenitor stars. Backing out population properties from this sample is difficult as the initial sample is seemingly heterogeneously selected.

\subsubsection{HST}\label{subsec:HST}
\citet{Ashley19} began with a parent sample of hot WDs showing near-infrared excess \citep{Farihi06,Farihi10}, supplemented by high-resolution imaging from the Hubble Space Telescope (HST), and performed follow-up spectroscopy to identify 16 WD + M dwarf PCEBs.
Their observed period distribution is roughly uniform in $\log (P_{\rm orb})$ from $0.1-10$ days, and because their RV detection efficiency is approximately constant across this range \citep{Ashley19}, this shape is not strongly biased by selection effects.
Although the sample is small, the results provide early, independent support for a log-uniform PCEB period distribution even out to longer periods ($\sim10$~days), consistent with the intrinsic distribution inferred in this work at $0.1<P_{\rm orb}/{\rm day}<2$ (Figure \ref{fig:period_dist}).
Furthermore, \citet{Ashley19} identify no binaries with orbital periods in the months-to-years regime, consistent with the expectation that most WD + M dwarf progenitors undergo unstable mass transfer and emerge from CEE at much shorter periods. For example, they find no binaries with periods longer than $10$~days, such as G 203-47-like binaries, despite being sensitive to (young) WDs in this region. This might suggest that G 203-47 being so close is somewhat fortunate.

\citet{Ashley19} also use the high-resolution HST images to find evidence for a second gap in the WDMS population between $10-100$~au.
They identify two distinct groups: close post-interaction binaries at $\lesssim 1-10$ au and wide binaries at $\sim100$-$1000$ au, with almost no systems in the intermediate $10-100$~au range. Whether such a gap is supported by current data remains unclear. {\it Gaia} DR3  wide binary catalogs show little support for such a discontinuity at $10-100$~au in the separation distribution of WD + M dwarf binaries \citep[e.g.,][]{EB18,EB21_widebin}, although current samples are small with the present {\it Gaia} angular resolution and sensitivity. Definitive tests will require the improved resolution and sensitivity anticipated from {\it Gaia} DR4.

At slightly larger secondary masses, \citet{Lagos22} surveyed WD + AFGK binaries with periods extending to $\sim100$ days, aiming to identify longer-period PCEBs ($10-50$~day periods). Their program is sensitive to systems with periods of weeks, yet all PCEBs they detect have $P_{\rm orb}<2.5$ days. The systems they do find at $10-40$ days are likely hierarchical triples with three MS stars rather than PCEBs. They interpret this as evidence for a transition in formation channels: very close WDMS binaries ($\lesssim 10$ days) arising from CEE, while somewhat wider WD+AFGK systems ($\gtrsim 100$ days) originate from non-conservative stable mass transfer. The latter population has since been better-characterized in the {\it Gaia} DR3 era and is consistent with arising from non-conservative stable mass transfer \citep[see][and references therein]{Yamaguchi25}.

\subsubsection{Gaia}\label{subsec:gaia}
Previous works have developed a large sample of unresolved WDMS binaries using the {\it Gaia} CMD \citep[e.g.,][]{RebassaMansergas21,RebassaMansergas25,Nayak25}. \citet{RebassaMansergas21} create a $100$~pc volume-limited sample from {\it Gaia} eDR3 and select all systems between the WD and MS tracks. Their CMD cuts are qualitatively similar to ours. They require that all systems in this volume satisfy the following astrometry and color quality filters: $\varpi/\sigma_\varpi\geq10$, $I_{\rm BP}/\sigma_{I_{\rm BP}}\geq10$, $\varpi/\sigma_\varpi\geq10$, 
$I_{\rm RP}/\sigma_{I_{\rm RP}}\geq10$, $I_{\rm G}/\sigma_{I_{\rm G}}\geq10$. These cuts are similar to those chosen in this work, but we also enforce that the systems be observable by ZTF ($\delta>-28^\circ$ and $G>12.5$) and additional fidelity/blending cuts are satisfied (${\tt fidelity\_v2}$ and ${\tt norm\_dG}$, see Section \ref{subsec:parent_sample}). After initial cuts, \citet{RM21} retain $2000$ WDMS candidates within $100$~pc, compared to our sample, which only contains $435$ within $100$~pc. They then remove systems that are either (1) more consistent with a single component SED, (2) blended with a nearby star, (3) a CV, or (4) have bad astrometry according to: ${\tt RUWE}>2$ or ${\tt astrometric\_excess\_noise}>2$ and ${\tt astrometric\_excess\_noise\_sig}>2$. This leaves $112$ WDMS candidates. Among these, $69$ are also in our parent sample. The remaining $43$ fail to enter our sample because $18$ are outside the ZTF footprint, and $25$ are just outside our CMD cuts (mostly too red).

\section{Connection to the broader WDMS population}\label{sec:comp_to_wideWDMS}

The binaries studied in this work are post--common-envelope and have short orbital periods ($\lsim2$~days). A complementary population of post--mass-transfer WDMS binaries has recently been identified at much wider separations ($P_{\rm orb} \sim 100-1000$~days) in the {\it Gaia} Non-Single Star (NSS) catalog \citep{Shahaf23,Shahaf24,Yamaguchi24}. They form the long-period counterparts to our close PCEBs and are an essential population for creating a coherent picture of binary mass transfer.
The WDMS binaries with $100-1000$~day orbital periods are likely products of {\it stable} mass transfer \citep{Hallakoun23,Garbutt24,Yamaguchi25,Shahaf25}\footnote{Common envelope evolution on the TP-AGB may also produce such wide systems \citep[e.g.,][]{Yamaguchi24,Belloni24b}}, in contrast to the \emph{unstable} common-envelope evolution that forms the short-period binaries studied here.

\subsection{Observed periods, masses, and distances}

Figure~\ref{fig:WDMS_corner} compares the orbital periods, masses\footnote{See Appendix \ref{app:SED_fitting} for details on deriving stellar parameters.}, and distances of our PCEB sample with the wider WDMS binaries from \citet{Yamaguchi24}, which are a spectroscopically confirmed subset of the \citet{Shahaf24} sample.
The {\it Gaia}-selected binaries typically contain solar-type ($0.7-1.1~{\rm M_\odot}$) MS companions as a result of the survey’s preference for bright, FGK-type secondaries \citep{Halbwachs23,EB24_gmock}. In contrast, our eclipsing PCEBs predominantly host lower-mass ($\lesssim0.4~M_\odot$) M dwarf companions. 
Both samples contain systems within only a few hundred parsecs and span a range of WD masses.

\begin{figure}
\centering
\includegraphics[width=0.99\columnwidth]{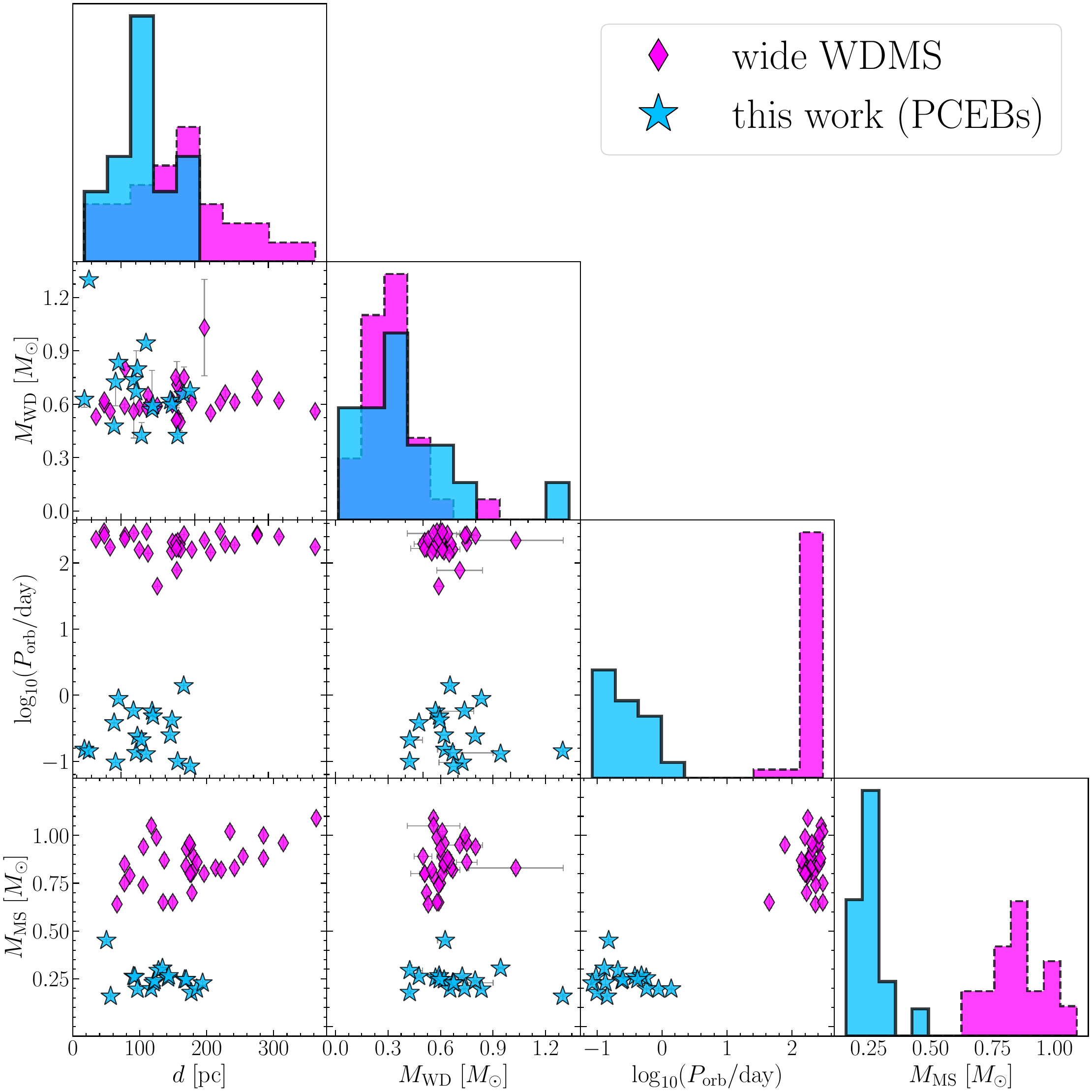}
\caption{
The distance ($d$), white dwarf mass ($M_{\rm WD}$), orbital period ($P_{\rm orb}$), and MS companion mass ($M_{\rm MS}$) of post--mass-transfer binaries. We compare wide WDMS binaries from {\it Gaia} NSS \citep[purple;][]{Yamaguchi24} to PCEBs identified in this work (blue). The diagonal histograms show the raw counts (not normalized). We only include PCEBs with GALEX UV photometry ($N=17$) because these systems have more reliable WD parameter estimates. Observed PCEBs in our sample cover shorter orbital periods and lower MS companion masses relative to the wider WDMS astrometric binaries.
}
\label{fig:WDMS_corner}
\end{figure}

The predicted space density of such wide WDMS binaries is $\approx2\times10^{-5}~{\rm pc^{-3}}$ for $100<P_{\rm orb}/{\rm day}<1000$, from which it is estimated that $0.4\%$ of solar-type stars have WD companions in this period range \citep{Yamaguchi25}. For the PCEBs in our work, we derive a slightly larger space density of $n = 7.2\times10^{-5}~{\rm pc^{-3}}$, suggesting that $\sim0.15\%$ of M dwarfs reside in a PCEB and that PCEBs are $\sim2-3\times$ more common than wide WDMS binaries.

One PCEB in our sample with GALEX FUV and NUV photometry has a best-fit WD mass of $M_{\rm WD}\approx1.3~{\rm M_\odot}$ ({\it Gaia} DR3 2273583445431091584). This newly-identified PCEB shows a peculiar light curve shape (Appendix \ref{app:all_LCs}) that could be consistent with a cyclotron-emitting magnetic WD in a near edge-on orbit \citep[e.g.,][]{vanRoestel25}. We note that the SED fit for this system, and therefore the derived stellar parameters, are less reliable given contamination from cyclotron emission in the optical.

\subsection{Intrinsic period distribution}

\begin{figure*}
    \centering
    \includegraphics[width=0.75\textwidth]{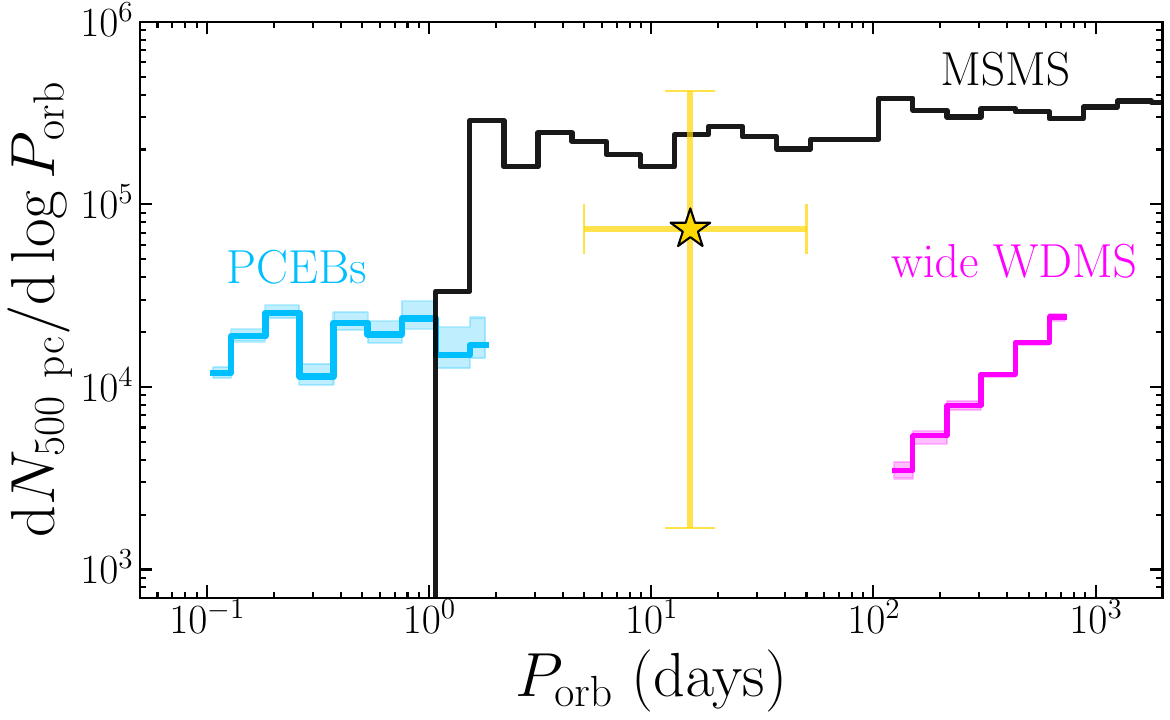}
    \caption{
    Period distribution of post--mass-transfer WDMS binaries within $500$~pc.
    We compare the inferred intrinsic period distribution of the WD + M dwarf PCEBs in our sample (blue, $0.1-2$~days) to the wider astrometric WD+FGK dwarf binaries from {\it Gaia} \citep[magenta, $100$-$1000$~days;][]{Yamaguchi25}. We also include the longer-period WD + M dwarf PCEB, G~203-47 (yellow star) with $P_{\rm orb}=14.9$~days at $d=7.5$~pc. We consider G~203-47 as the only binary of its class (e.g., $P_{\rm orb} \in 5-50$~days) within $10$~pc; $2\sigma$ Poisson error bars ($N=1$) are displayed for this binary.
    The PCEBs are roughly flat in $\log P_{\rm orb}$, whereas the wide post-interaction WDMS binaries are roughly flat in $P_{\rm orb}$. For reference, we show the period distribution of all main-sequence binaries with $M_1>1~{\rm M_\odot}$ (black). PCEBs are $2-3\times$ more common than wide WDMS binaries, but binaries between these two regimes ($\sim5$-$100$~days) remain poorly constrained.}
    \label{fig:all_period_dist}
\end{figure*}

To gain a more holistic view of the post--mass-transfer binary population, we compare the inferred period distributions for different classes of post-interaction WDMS binaries within $500$~pc. In particular, we define ${\rm d}N_{\rm 500~pc}/{\rm d}\log P_{\rm orb}$ as the mean number of binaries per logarithmic interval of $P_{\rm orb}$. Given $\Delta N_{\rm 500~pc}$, the intrinsic number of binaries within $500$~pc in a given logarithmic period bin of width $\Delta \log P_{\rm orb}$, we define
\begin{equation}
    \frac{{\rm d}N_{\rm 500~pc}}{{\rm d}\log P_{\rm orb}} = \frac{\Delta N_{\rm 500~pc}}{\Delta \log P_{\rm orb}}.
\end{equation}
Integrating this curve over a given period range for a specific binary class will recover the total number of binaries within $500$~pc in that class and period range. The considered binary classes are discussed below.

\citet{Yamaguchi25} modeled the {\it Gaia} NSS selection function \citep[see][]{EB24_gmock} and recovered an intrinsic period distribution for wide WDMS binaries that is approximately uniform in $P_{\rm orb}$ over $100$–$1000$~days; specifically, they find ${\rm d}N/{\rm d}P_{\rm orb} \propto P_{\rm orb}^{\beta}$ with $\beta = 0.12 \pm 0.08$. By forward-modeling their selection function, they also conclude that $10{,}758$ such systems reside within $500$~pc of the Sun. 
To generate a corresponding mock population within $500$~pc, we draw $\beta$ from the corresponding Gaussian distribution and sample $10{,}758$ orbital periods from the resulting $P_{\rm orb}^{\beta}$ distribution over $100<P_{\rm orb}/{\rm day}<1000$.
We repeat this procedure 100 times to produce a median period distribution with systematic uncertainties for wide WDMS binaries, which we then compare directly to the PCEBs within the same volume.

The period distribution of our PCEB sample is already corrected for the eclipse-detection incompleteness (Figure~\ref{fig:period_dist}). We further correct for the CMD-incompleteness by scaling the corrected histogram by a factor of $7.2$ to reflect that only $21\%$ of PCEBs reside in our CMD-selected region and only $\approx66\%$ are in the ZTF footprint ($\delta>-28^\circ$). Together, these factors yield the full completeness correction for the $200$~pc volume-limited sample, which is scaled by $\tilde{V}(500)/\tilde{V}(200)=11.5$ to estimate the population within $500$~pc.
Applying these corrections predicts a total $\approx 25{,}000$ PCEBs within $500$~pc, $2-3\times$ larger than the predicted wide post-interaction WDMS population of $\approx 11{,}000$ \citep[][]{Yamaguchi25}. 
At present, the observed wide WDMS period distribution is truncated at $\sim1000$~days, primarily set by the {\it Gaia} DR3 baseline. Because their period distribution rises toward longer periods, the true number of wide post–interaction WDMS binaries is expected to increase as longer-period binaries become accessible in future {\it Gaia} releases.
The total expected number of MSMS binaries within $500$~pc is $\sim8.4\times10^{6}$.

We also include the longer-period PCEB G 203-47 \citep{Delfosse99} discussed in Section \ref{subsec:SDSS}. 
To estimate the space density of intermediate-period post-interaction binaries (i.e., ``G 203-47-like''), we assume there is $1$ within $10$~pc in the period range of $5-50$ days; the original RV survey was complete to M dwarfs within $9$~pc and found only one with a WD companion \citep{Delfosse98,Delfosse99}. The $10$~pc space density is then scaled to $500$~pc ($\tilde{V}(500)/\tilde{V}(10)$) with $2\sigma$ Poisson confidence intervals propagated. We note that G 203-47 may be fortuitously close, given that other searches sensitive to young WD + M dwarf binaries at $P_{\rm orb}\approx15$~days recover no such systems \citep[][see also the discussion in Section \ref{subsec:HST}]{Ashley19}.

Figure~\ref{fig:all_period_dist} presents the resulting period distributions for both populations of post--mass-transfer WDMS binaries within $500$~pc. 
For comparison, we include the underlying distribution of all non-interacting MSMS binaries with $M_1 \geq 1~{\rm M_\odot}$ (i.e., possible WD progenitors) following the distributions of \citet{Moe17}. The total expected number of such MSMS binaries within $500$~pc is $\sim2\times10^{6}$.

As Figure~\ref{fig:all_period_dist} illustrates, PCEBs exhibit a log-flat distribution (${\rm d}N/{\rm d}\log P_{{\rm orb}}\propto P_{{\rm orb}}^{0}$) at $P_{\rm orb} < 2$ days whereas the wide post-interaction WDMS are log-increasing (${\rm d}N/{\rm d}\log P_{{\rm orb}}\propto P_{{\rm orb}}^{1}$) at $P_{\rm orb} = 100-1000$ days.
The two populations occupy distinct regions of period space and likely arise from different evolutionary channels: wide WDMS binaries are mainly produced via {\it stable} mass transfer \citep[e.g.,][]{Garbutt24,Yamaguchi25,Shahaf25}, while the short-period PCEBs emerge from {\it unstable} common-envelope evolution.
These two populations also have different secondary mass distributions: the MS stars in our PCEB sample are mainly M dwarfs, while those in the wide WDMS sample are mainly FGK dwarfs.

The gap within $10-100$~days is mainly due to observational limitations: at these periods, WDMS do not show significant photometric variability (neither eclipses, reflection, or ellipsoidal), they are rarely identified by few-epoch RV campaigns, and they do not present a strong enough astrometric signal to obtain a {\it Gaia} DR3 orbital solution. Observational efforts to empirically constrain the period distribution in this regime are needed.

Together, the short- and long-period post-interaction WDMS populations provide complementary constraints on the stability of mass transfer and the bifurcation between the stable and unstable regimes. Future work constraining the intrinsic period distribution across the intermediate range ($P_{\rm orb}\in 10-100$~days) will be essential for building a holistic picture of binary mass transfer.

\section{Conclusions}\label{sec:conclusions}
In this work, we create a volume-limited sample of $\sim3800$ unresolved WDMS binary candidates within $200$~pc by selecting systems between the WD and MS tracks in the {\it Gaia} CMD (Figure \ref{fig:gaia_cuts}).
From this parent sample, we search for WDMS eclipsing binaries using ZTF light curves (e.g., Figure \ref{fig:example_LCs}). We identify $39$ detached eclipsing PCEBs with $P_{\rm orb} \in 0.1-2$~days with cool M dwarf companions ($M_{\rm MS}\lsim0.4~{\rm M_\odot}$) and relatively hot (young) WDs ($T_{\rm eff,WD}\gsim9{,}000$~K). Most importantly, our methodology explicitly maintains well-defined selection criteria at each step, enabling a clean and interpretable sample. 
We model our survey's selection function (e.g., Figure \ref{fig:completeness} and Figure \ref{fig:completeness_mass}) to correct the observed sample for incompleteness and infer {\it intrinsic} demographics of the PCEB population. Our main results are summarized as follows:

\begin{enumerate}
    \item {\it Orbital period distribution:} PCEBs show an intrinsic orbital period distribution consistent with being uniform in $\log(P_{\rm orb})$ between $0.1$ and at least $2$~days (Figure \ref{fig:period_dist}). Combining our results with spectroscopic surveys \citep[e.g.,][]{Ashley19} suggests that the period distribution remains log-uniform out to at least $10$~days. 

    \item {\it Companion mass distribution:} The intrinsic companion-mass distribution of short-period WD + M dwarf PCEBs peaks around $M_{\rm MS}\simeq0.25~{\rm M_\odot}$ and declines steeply toward higher masses in the range $M_{\rm MS} = 0.25-0.5~{\rm M_\odot}$ (Figure \ref{fig:mass_dist}). 

    \item {\it Space density and birth rate:} Our inferred space density of WD + M dwarf binaries with $P_{\rm orb} < 2$~days is $n = 7.2\times10^{-5}~{\rm pc^{-3}}$, suggesting that $1$ in $\sim60$ white dwarfs and $1$ in $\sim650$ M dwarfs reside in short-period PCEBs. The birth rate ($0.01~{\rm yr^{-1}}$) implies that a new WD + M dwarf binary emerges out of common envelope roughly every century in the Galaxy. Both the space density and Galactic birth rate provide key anchors for binary population models. 

    \item {\it Implications for magnetic braking and CV formation:} Forward evolution suggests that roughly half of newly formed PCEBs would reach Roche-lobe overflow within a Hubble time if magnetic braking operates efficiently in fully-convective stars. However, this overpredicts the observed CV birth rate by a factor of $\sim10$, favoring models where magnetic braking is disrupted below the fully convective boundary or where the stars merge instead of becoming long-lived CVs. The fact that $38/39$ of the PCEBs in our sample have companion masses below the fully convective boundary, despite our search being sensitive to higher-mass companions, seems to provide additional support for disrupted magnetic braking.
    
    \item {\it A global view of post-interaction WDMS binaries:} 
    By combining our PCEB sample with constraints on wider WDMS binaries, we construct an empirical picture of post-interaction WDMS binaries spanning orbital periods from $0.1$ to $1000$~days. The emerging period distribution is roughly log-flat (${\rm d}N/{\rm d}\log P_{{\rm orb}}\propto P_{{\rm orb}}^{0}$) at $P_{\rm orb} < 2$ days and log-increasing (${\rm d}N/{\rm d}\log P_{{\rm orb}}\propto P_{{\rm orb}}^{1}$) at $P_{\rm orb} = 100-1000$ days. While the $10-100$ day regime is poorly constrained by existing surveys, a few nearby systems suggest that it is also well-populated. 
    Short-period PCEBs with M dwarf companions are $\sim2-3$ times more common than wider post-interaction WDMS binaries with FGK companions, which likely formed through stable mass transfer.
\end{enumerate}

Constraining the demographics (periods, masses, rates) of post-interaction WDMS binaries across a wider range of periods and companion masses is essential for studying the full landscape of binary mass transfer. The well-characterized PCEB sample presented here provides a robust anchor at short periods and low companion masses, where selection effects are explicitly quantified and corrected. However, populations of short-period PCEBs containing FGK stars and long-period WD + M dwarf binaries remain observationally unconstrained.
Future time-domain surveys, deeper RV monitoring, and improved {\it Gaia} astrometry will be essential for filling in these gaps and linking the short-period PCEBs presented here to the full population of post-interaction WDMS binaries. Ultimately, mapping out this landscape will provide sharper empirical constraints on mass transfer stability, common-envelope evolution, and angular momentum loss in binary star systems.

\section{Acknowledgments} \label{acknowledgments}
We thank the anonymous referee for constructive comments that helped improve the manuscript. We are grateful for the KITP Stellar-Mass Black Holes Workshop for inspiring conversations and Sahar Shahaf, Natsuko Yamaguchi, Dominick Rowan, Hans-Walter Rix, Samuel Whitebook, and Matteo Cantiello for useful discussions.

C.S. acknowledges support from the Department of Energy Computational Science Graduate Fellowship.
This material is based upon work supported by the U.S. Department of Energy, Office of Science, Office of Advanced Scientific Computing Research, under Award Number DE-SC0026073. This research was supported by NSF grants AST-2307232 and AST-2508988. This research was supported in part by grant NSF PHY-2309135 to the Kavli Institute for Theoretical Physics (KITP). 

This work presents results from the European Space Agency (ESA) space mission {\it Gaia}. {\it Gaia} data are being processed by the {\it Gaia} Data Processing and Analysis Consortium (DPAC). Funding for the DPAC is provided by national institutions, in particular the institutions participating in the {\it Gaia} MultiLateral Agreement (MLA). The {\it Gaia} mission website is \url{https://www.cosmos.esa.int/Gaia}. The {\it Gaia} archive website is \url{https://archives.esac.esa.int/Gaia}.

Based on observations obtained with the Samuel Oschin Telescope 48-inch and the 60-inch Telescope at the Palomar Observatory as part of the Zwicky Transient Facility project. ZTF is supported by the National Science Foundation under Grant No. AST-2034437 and a collaboration including Caltech, IPAC, the Weizmann Institute for Science, the Oskar Klein Center at Stockholm University, the University of Maryland, Deutsches Elektronen-Synchrotron and Humboldt University, the TANGO Consortium of Taiwan, the University of Wisconsin at Milwaukee, Trinity College Dublin, Lawrence Livermore National Laboratories, and IN2P3, France. Operations are conducted by COO, IPAC, and UW. The ZTF forced-photometry service was funded under the Heising-Simons Foundation grant $\#12540303$ (PI: Graham).

\software{This work made use of the following software packages: \texttt{astropy} \citep{astropy:2013,astropy:2018,astropy:2022}, \texttt{Jupyter} \citep{2007CSE.....9c..21P,kluyver2016jupyter}, \texttt{matplotlib} \citep{Hunter:2007}, \texttt{numpy} \citep{numpy}, \texttt{pandas} \citep{mckinney-proc-scipy-2010,pandas_17806077}, \texttt{python} \citep{python}, \texttt{scipy} \citep{2020SciPy-NMeth,scipy_17467817}, \texttt{COSMIC} \citep{COSMIC,COSMIC_13351205}, \texttt{Cython} \citep{cython:2011}, \texttt{emcee} \citep{emcee-Foreman-Mackey-2013,emcee_10996751}, \texttt{h5py} \citep{collette_python_hdf5_2014,h5py_7560547}, \texttt{schwimmbad} \citep{schwimmbad}, \texttt{seaborn} \citep{Waskom2021}, \texttt{TOPCAT} \citep{2005ASPC..347...29T}, and \texttt{tqdm} \citep{tqdm_14231923}.
Software citation information aggregated using \texttt{\href{https://www.tomwagg.com/software-citation-station/}{The Software Citation Station}} \citep{software-citation-station-paper,software-citation-station-zenodo}.}
\appendix 

\section{Sensitivity of our PCEB search}\label{app:sensitivity}

To quantify the regions of parameter space to which our PCEB search is sensitive, we use our synthetic {\tt COSMIC} population of WDMS binaries as described in Section~\ref{subsubsec:parent_sample_completeness}. For this specific experiment, we apply a ${\tt q_{\rm crit}=0.4}$ following \citet{Yamaguchi25}, such that mass transfer at $q<q_{\rm crit}$ is unstable, while the rest is stable. The implications of this choice are discussed further later in this section.
Next, we select systems within $200$~pc, restrict to detached PCEBs in the period range, and compute synthetic {\it Gaia} photometry and absolute magnitudes for each binary.

We then apply the same CMD selection criteria and photometric-quality cuts used in our PCEB search (Section~\ref{subsec:parent_sample}). Systems that satisfy all these observational cuts are classified as ``in region'' and represent the simulated population that would be detectable by our procedure. The remaining systems form the broader ``all simulated PCEBs'' distribution. We make no claim that the simulated sample's demographics represent the Solar neighborhood PCEB population, but instead use their broad range of properties to understand the landscape of WD and MS parameters that our {\it Gaia} search is sensitive to.

\begin{figure*}[h]
\centering
\includegraphics[width=0.95\columnwidth]{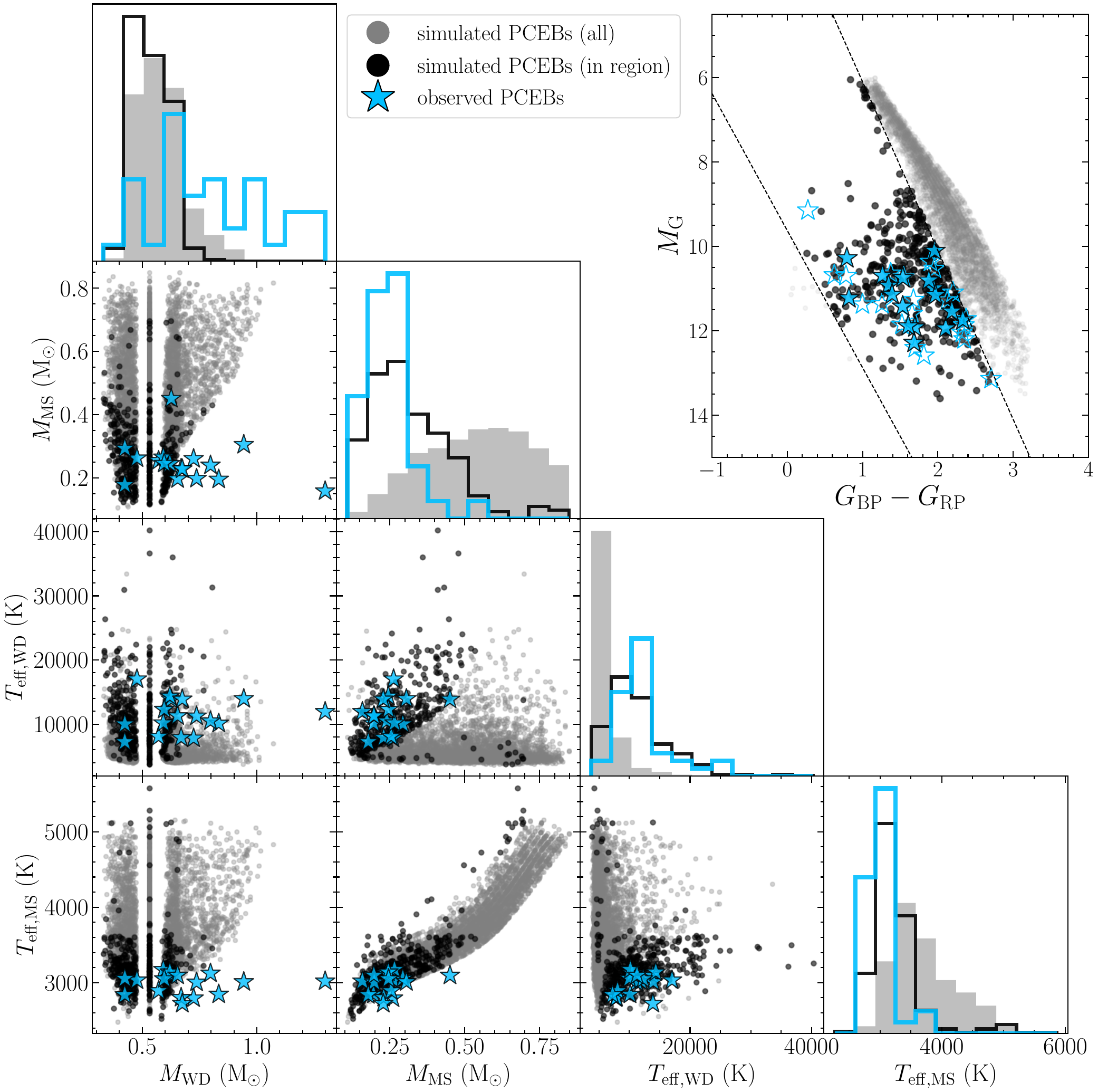}
\caption{
Sensitivity of our PCEB search. The corner plot compares simulated PCEBs (gray), simulated systems that fall within our CMD-selected region (black), and the final observed PCEB sample (blue). Diagonal panels show the one-dimensional distributions of WD mass, MS mass, WD effective temperature, and MS effective temperature. Off–diagonal panels show the corresponding scatter plots. Gray points indicate the full {\tt COSMIC} population of detached PCEBs within 200 pc satisfying the initial {\it Gaia} query (Section \ref{subsec:parent_sample}), while black points highlight those that also reside in the CMD-selected region. Blue stars show the measured parameters of the PCEBs recovered in this work with {\it GALEX} UV photometry (hollow stars in the CMD  are without UV photometry). The top-right panel shows the location of these systems in the {\it Gaia} CMD relative to the full simulated WDMS population, illustrating how our selection preferentially recovers binaries containing hotter WDs and cooler, lower-mass secondaries.}
\label{fig:sensitivity_corner}
\end{figure*}

Figure~\ref{fig:sensitivity_corner} summarizes the sensitivity of our search. The gray points show the full {\tt COSMIC} population of detached PCEBs within $200$~pc, while black points highlight those systems that fall within our observational selection region. Blue stars show the 39 PCEBs identified in this work. The diagonal panels compare 1D distributions of mass and effective temperatures for both components; the off-diagonal panels show the corresponding 2D scatter plots. The discrete feature at $M_{\rm WD}=0.55$ represents an artifact of {\tt COSMIC}'s WD initial-final mass relation. The WD effective temperature for the synthetic population, $T_{\rm eff, WD}$, is calculated by interpolating MIST models \citep{Bauer25} using the reported WD mass and cooling age from {\tt COSMIC}. 

These comparisons illustrate several key features of our search sensitivity. First, our CMD cuts efficiently reject systems with bright, relatively massive MS companions, leading to a strong bias toward binaries with M dwarf secondaries ($M_{\rm MS}\lesssim0.5~{\rm M_\odot}$). Second, our requirement that systems fall in the {\it Gaia} CMD gap selects preferentially for binaries containing hot ($T_{\rm eff,WD}\gtrsim10{,}000$ K) WDs, for which the WD contributes a significant fraction of the optical flux. Cooler WDs in binaries with mid-M companions become difficult to distinguish from the main sequence and are rarely recovered. 

Interestingly, this exercise reveals that our search is sensitive to PCEBs with companions up to $M_{\rm MS}=0.5~{\rm M_\odot}$, despite there being only $4/39$ PCEBs in our observed sample having $M_{\rm MS}>0.3~{\rm M_\odot}$ and only one with $M_{\rm MS}>0.35~{\rm M_\odot}$.
For example, the synthetic {\it Gaia} CMD in Figure \ref{fig:sensitivity_corner} reveals a population of PCEBs at ($G_{\rm BP}-G_{\rm RP}$, $M_{\rm G}$) = ($2$,$10$), corresponding to companion masses of $0.3-0.45~{\rm M_\odot}$, that is not observed in our PCEB sample despite being sensitive there. 
This discrepancy cannot be attributed to selection effects alone, plausibly reflecting an underlying feature of PCEB formation and/or evolution. 

One possibility is that the effective stability boundary for mass transfer during the CEE phase lies near $q_{\rm crit}\simeq0.3-0.4$, naturally favoring post--CEE binaries with low-mass secondaries \citep[][]{Yamaguchi25}. Alternatively, a broad range of companion masses may emerge from the CE, but systems with partially radiative secondaries ($M_{\rm MS}\gtrsim0.35~{\rm M_\odot}$) may evolve rapidly out of the detached PCEB phase if MB is stronger above the fully convective limit, as predicted by disrupted and boosted MB models. The sharp decline in the companion-mass distribution above $M_{\rm MS}\approx0.25-0.30~{\rm M_\odot}$ of our observed sample appears somewhat consistent with this interpretation, although the cutoff is lower than the nominal fully convective boundary ($0.35~{\rm M_\odot}$). Furthermore, the stark dearth of $M_{\rm MS}>0.35~{\rm M_\odot}$ systems would require a large boosting factor for the disrupted MB model. Future work is needed to discriminate between these two scenarios.

The WD mass distributions also differ noticeably between the simulated and observed samples. While the \texttt{COSMIC} model peaks near $0.5-0.6~{\rm M_\odot}$, the observations show a broader distribution with a significant tail toward higher WD masses. The photometric WD masses shown here are derived from SED fitting assuming mass–radius relation (Appendix \ref{app:SED_fitting}), and thus carry systematic and model uncertainties. However, systems with UV photometry, where the WD radius is slightly better constrained, also favor relatively high WD masses (Figure \ref{fig:WDMS_corner}), suggesting that this trend might be partly physical rather than purely an artifact of the modeling. A more detailed analysis of the WD mass distribution, including UV spectroscopy, will be needed to confirm and interpret these differences.

Overall, comparing PCEBs inside the CMD region to the overall population clarifies the ranges of MS and WD parameters that our search is most sensitive to: hot WDs $T_{\rm eff,WD}\gsim8000~{\rm K}$ and cool, low-mass MS companions ($M_{\rm MS}\lsim0.45~{\rm M_\odot}$). We consider the sensitivity limits when interpreting the demographics of our PCEB sample and comparing to theoretical population models.

\section{ZTF light curves}\label{app:all_LCs}
\begin{figure*}
\centering
\includegraphics[width=1.0\columnwidth]{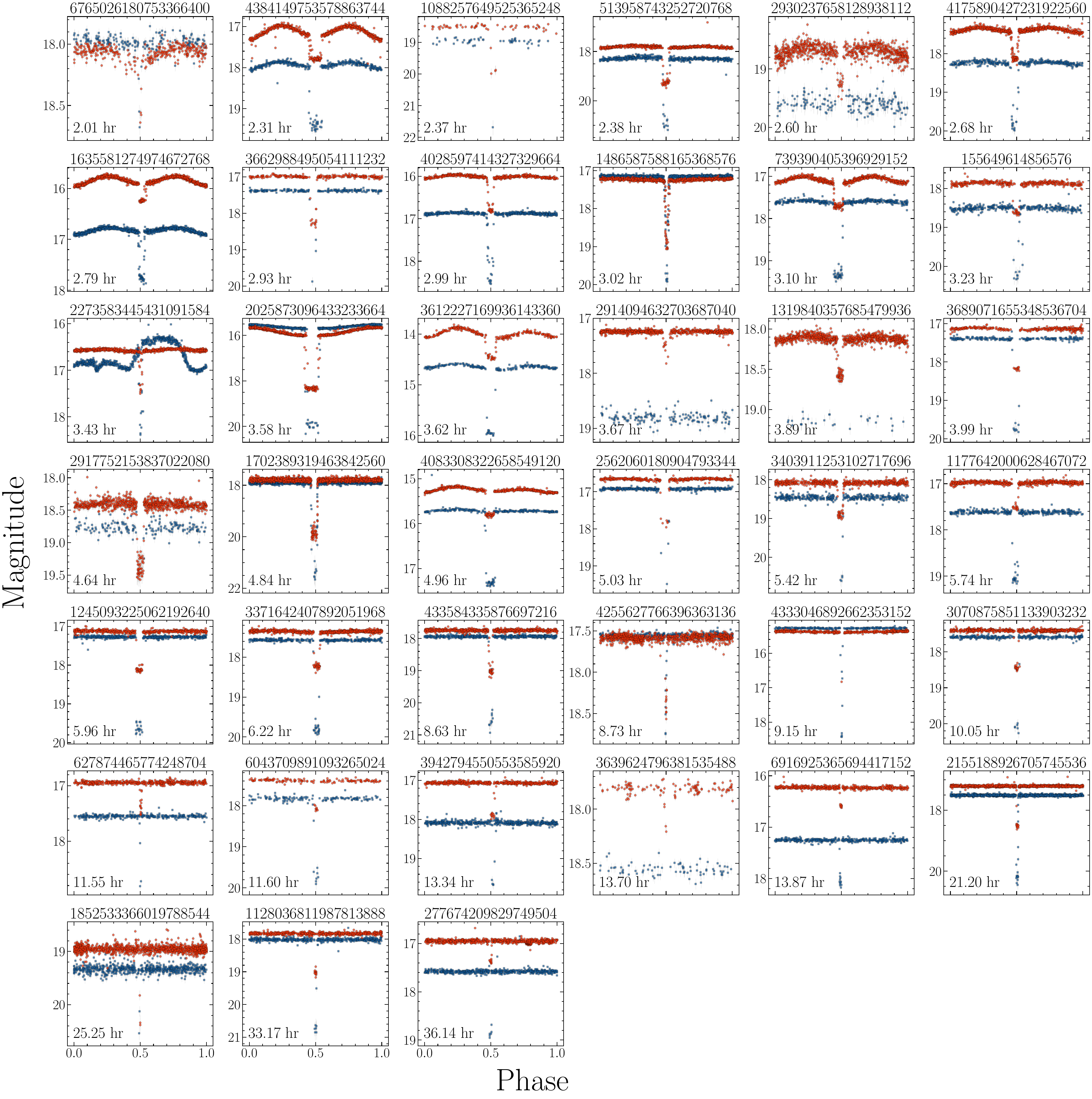}
\caption{
Light curves for all PCEBs in our sample. The red circles (blue squares) show the measured photometry in the ZTF $r$-band ($g$-band). Each light curve is folded on its best-fit period. 
}
\label{fig:WDMS_phasefold_grid_gr}
\end{figure*}

In Figure \ref{fig:WDMS_phasefold_grid_gr} we show all ZTF light curves in $g$- and $r$-bands for the white dwarf MS eclipsing binaries in our sample. Each light curve is folded on the best-fit period with its {\it Gaia} DR3 Source ID above the light curve. The phase is shifted such that the primary eclipse occurs around $0.5$.

\section{Spectral Energy Distribution Fitting}\label{app:SED_fitting}

We assembled available broadband photometry (in Vega mag) for each source from \textit{Gaia} DR3 \citep[$G$, $B_P$, $R_P$;][]{GaiaDR3}, GALEX \citep[$FUV$, $NUV$;][]{Martin05}, SDSS DR16 \citep[$ugrizy$;][]{SDSS_DR16}, 2MASS \citep[$J$, $H$, $K$;][]{Skrutskie06}, and \textit{WISE} \citep[$W1,W2,W3,W4$;][]{Wright10,Cutri14}. The spectral energy distributions (SEDs) were modeled as a combination of two components: a MS star and a WD. For the MS component, we interpolated model spectra from the \texttt{BT-Settl} library \citep{Allard13} using the \texttt{pystellibs}\footnote{\url{https://github.com/mfouesneau/pystellibs}} package. The WD fluxes were computed from the Koester model grids \citep{Koester10}. The total model spectrum is the sum of both components, reddened by a wavelength-dependent extinction curve following \citet{Cardelli89}. Synthetic photometry is derived by integrating the combined flux through each observed filter response using \texttt{pyphot} \citep{zenodopyphot}.

We fit each SED using a Markov Chain Monte Carlo (MCMC) approach implemented using \texttt{emcee} \citep{emcee-Foreman-Mackey-2013}. The $4$ free parameters include the effective temperatures ($T_{\mathrm{eff}}$) and radii ($R$) of both MS and WD components. We fix the line-of-sight extinction $E(B-V)$ to the value derived from the \citet{Edenhofer24} 3D dust map and calculate the mass of each component using mass-radius relations of \citet{Baraffe15} for MS stars and \citet{Bedard17} for WDs. As a result, the reported mass uncertainties in Table \ref{tab:WDMS_params} are underestimated because they do not consider model dependencies. Empirical measurements indicate that M dwarfs are often inflated by $\sim$5–15\% relative to model predictions \citep[e.g.,][]{LopezMorales05,Parsons18,Kesseli18,Brown22}, which would bias our inferred masses high by a comparable fractional amount. We also test an updated distribution $M-R$ relation from \citet{Brown22}, similar to observations from \citet{Parsons18}, finding that it increases the number of systems in the lowest mass bin ($\approx0.1~{\rm M_\odot}$), but it does not change the shape of the intrinsic distribution significantly since this bin also has the largest detection probability (i.e., most complete).
Lastly, we note that the systematic mass uncertainty between different models is smaller than the width of the companion-mass distribution and does not affect our qualitative conclusions regarding the dominance of low-mass companions in the PCEB population.

Uniform priors were adopted for all fitted parameters except for the WD radius ($R_{\mathrm{WD}}$). 
For $R_{\rm WD}$, we instead adopt a physically motivated prior corresponding to a Gaussian on the WD mass, with mean $M_{\rm WD}=0.5~{\rm M_\odot}$ and standard deviation $\sigma=0.1~{\rm M_\odot}$, mapped to radius via a standard WD mass–radius relation (see below). This choice is motivated by empirical constraints from the SDSS PCEB population, which shows that PCEBs preferentially host relatively low-mass white dwarfs \citep[e.g.,][]{Zorotovic11a}. In addition, we discovered that adopting a uniform prior in $R_{\rm WD}$ (or $M_{\rm WD}$) systematically biases the fits toward larger WD masses, whereas the Gaussian mass prior yields more physically plausible solutions.
The adopted prior ranges were:
$T_{\mathrm{eff,MS}} \in [2000, 5000]$~K, $R_{\mathrm{MS}} \in [0.1, 1.0]~R_\odot$, $T_{\mathrm{eff,WD}} \in [5000, 45000]$~K, $R_{\mathrm{WD}} \in [0.005, 0.05]~R_\odot$.
Each fit is initialized with 100 walkers evolved for 150 burn-in steps, followed by 200 production steps to sample the posterior distributions. The median posterior values were adopted as the best-fit parameters, and the 16th-84th percentiles as their corresponding $1\sigma$ uncertainties. 
Table \ref{tab:WDMS_params} provides the full table of parameters and their uncertainties\footnote{A machine-readable version of Table~\ref{tab:WDMS_params}, including an expanded set of derived and observed parameters for each source, is available at \url{https://github.com/cheyanneshariat/pcebs}.}, and Figure \ref{fig:example_SEDfit} the SED results for our two-component fits. For WD analyses in this paper, we focus on the sources with GALEX UV photometry, such that the WD parameters are somewhat well-constrained by the SED, although UV spectra are undoubtedly required for accurate parameter estimation. As such, we do not report masses for WDs without UV photometry. The WD cooling age is derived using MIST models \citep{Bauer25}, which only provides tracks for a minimum WD mass of $\sim0.5~{\rm M_\odot}$. Cooling ages for WDs below this minimum mass are calculated assuming $M_{\rm WD}=0.5~{\rm M_\odot}$.

We identify systems that are blended with nearby sources (Gaia DR3 562060180904793344 and 277674209829749504), as well as one with previously-identified IR excess \citep[Gaia DR3 155649614856576;][]{Debes12,Parsons13b}. For this particular system, the excess arises from cyclotron emission due to the $8$~BG magnetic field \citep{Parsons13b}. We fit the SED without the near-IR photometry, finding a reasonably good fit for the UV-optical (Figure \ref{fig:example_SEDfit}) and parameters in agreement with \citet{Parsons13b}. Another PCEB in our sample, {\it Gaia} DR3 2273583445431091584, shows significant optical cyclotron emission (Figure \ref{fig:all_SEDs_2}), which may also alter its SED fit. This is the only PCEB with such a light curve in our sample.
While undetected cyclotron emission (particularly in the NIR) may affect SED fits of magnetic systems, such objects are expected to be uncommon in our sample. Magnetic WDs are a minority among WDs and preferentially arise in more massive WDs \citep[e.g.,][]{Liebert88,Dobbie12,Dobbie13}, whereas PCEBs preferentially contain low-mass WDs \citep[e.g.,][]{Zorotovic10}. This makes magnetic systems less likely to enter our CMD-selected WDMS sample, suggesting that cyclotron contamination should be minimal and not impact our main results that depend on SED-inferred parameters.

The photometry entering the SED fits is obtained at a range of orbital phases, and in eclipsing systems, the instantaneous flux can deviate from the out-of-eclipse level. However, because the eclipse duty cycle is small and the broadband measurements (e.g., {\it Gaia}, 2MASS, GALEX) are typically averaged over multiple epochs or obtained sparsely in time, the SEDs are expected to be dominated by out-of-eclipse flux. While phase-dependent variability may increase the scatter in individual fits, it is not expected to introduce a systematic bias in the population-level temperature and radius estimates presented here, particularly due to the short duty-cycle of WD eclipses.

\section{Magnetic Braking Equations}\label{app:MB_equations}

Here we summarize the magnetic braking prescriptions adopted in this work and the corresponding analytic solutions for orbital period evolution. These equations are used to evolve detached PCEBs forward in time (until Roche lobe crossing) and to estimate the delay between common-envelope ejection and the onset of Roche--lobe overflow. Note that gravitational wave radiation is almost always negligible compared to MB
at all periods where PCEBs are detached \citep[e.g.,][]{Peters64}.

We assume circular, tidally synchronized binaries throughout, such that the stellar rotation period equals the orbital period, $P_{\rm rot} = P_{\rm orb}$.
The orbital angular momentum of a circular binary is
\begin{equation}
J_{\rm orb}
=
\frac{G^{2/3}}{(2\pi)^{1/3}}
\frac{M_{\rm WD} M_{\rm MS}}{(M_{\rm WD}+M_{\rm MS})^{1/3}}
P_{\rm orb}^{1/3},
\end{equation}
where $M_{\rm WD}$ and $M_{\rm MS}$ are the white dwarf and main-sequence star masses and $P_{\rm orb}$ is the orbital period.
Angular momentum losses due to magnetic braking remove angular momentum from the orbit,
$\dot{J}_{\rm orb} = \dot{J}_{\rm MB}$, leading to orbital period evolution
$
\dot{P}_{\rm orb}
=
3~P_{\rm orb}~(\dot{J}_{\rm orb}/J_{\rm orb}).
$

\subsection{Classical (RVJ) magnetic braking}

We adopt the classical magnetic braking prescription of \citet{Verbunt81,Rappaport83}, motivated by the Skumanich spin-down law. The magnetic braking torque acting on a single star is
\begin{equation}
\dot{J}_{\rm RVJ}
=
-\tau_0
\left(\frac{M}{M_\odot}\right)
\left(\frac{R}{R_\odot}\right)^{4}
\left(\frac{P_{\rm rot}}{1~{\rm d}}\right)^{-3},
\end{equation}
where $M$ and $R$ are the stellar mass and radius, and $\tau_0 \simeq 6.8\times10^{34}$ erg sets the normalization.

For a tidally synchronized binary, this torque yields an analytic solution for the orbital period evolution,
\begin{equation}
P_{\rm orb}(t)
=
\left[
P_{\rm orb,0}^{10/3}
-
\frac{10\,R_{\rm MS}^4\,(q+1)^{1/3}}
{q\,M_{\rm MS}^{2/3}}
\,t
\right]^{3/10},
\end{equation}
where $P_{\rm orb,0}$ is the orbital period immediately after common-envelope ejection, $M_{\rm MS}$ and $R_{\rm MS}$ are the companion mass and radius in solar units, and $q = M_{\rm WD}/M_{\rm MS}$.
The dimensionless time is defined as
$
t \equiv T/T_{\rm RVJ},
$
with the RVJ timescale
$
T_{\rm RVJ}
=
(1~{\rm d})^{1/3}
M_\odot^{5/3}
\left(\frac{G^2}{2\pi}\right)^{1/3}
\tau_0^{-1}
\approx 5.8~{\rm Gyr}.
$

\subsection{Saturated magnetic braking}

We also consider a saturated magnetic braking prescription following \citet{Sills00}, motivated by observations that magnetic activity saturates above a critical rotation rate. The magnetic braking torque on a single star is
\begin{equation}
\dot{J}_{\rm sat}
=
\begin{cases}
-\tau_1
\left(\dfrac{P_{\rm rot}}{1~{\rm d}}\right)^{-3}
\left(\dfrac{R}{R_\odot}\right)^{1/2}
\left(\dfrac{M}{M_\odot}\right)^{-1/2},
& P_{\rm rot} \ge P_{\rm crit}, \\
\\
-\tau_1
\left(\dfrac{P_{\rm rot}}{1~{\rm d}}\right)^{-1}
\left(\dfrac{P_{\rm crit}}{1~{\rm d}}\right)^{-2}
\left(\dfrac{R}{R_\odot}\right)^{1/2}
\left(\dfrac{M}{M_\odot}\right)^{-1/2},
& P_{\rm rot} < P_{\rm crit},
\end{cases}
\end{equation}
where $\tau_1 = 1.04\times10^{35}$ erg and the saturation threshold $P_{\rm crit} = 0.1~P_{\odot}(\tau_c/\tau_{c,\odot})$ depends on the convective turnover timescale, $\tau_c$ \citep{Wright11}.

In the saturated regime relevant for most systems in our sample, the analytic solution for the orbital period evolution is
\begin{equation}
P_{\rm orb}(t)
=
\left[
P_{\rm orb,0}^{4/3}
-
\frac{4\,R_{\rm MS}^{1/2}\,(q+1)^{1/3}}
{q\,M_{\rm MS}^{13/6}}
\,t
\right]^{3/4}.
\end{equation}
The dimensionless time is defined as
$
t \equiv \frac{T}{T_{\rm sat}},
$
where the saturated magnetic braking timescale is
\begin{equation}
T_{\rm sat}
=
\frac{
(1~{\rm d})^{4/3}
M_\odot^{5/3}
G^{2/3}
}{
(2\pi)^{4/3}
K_W
\omega_{\rm crit}^2
}.
\end{equation}
Here $\omega_{\rm crit,1} = 2\pi / P_{\rm crit}$, and
$K_W = 2.7\times10^{47}~{\rm g~cm^2~s}$ is the magnetic braking constant from \citet{Sills00}.

\begin{figure}
\centering
\includegraphics[width=0.85\columnwidth]{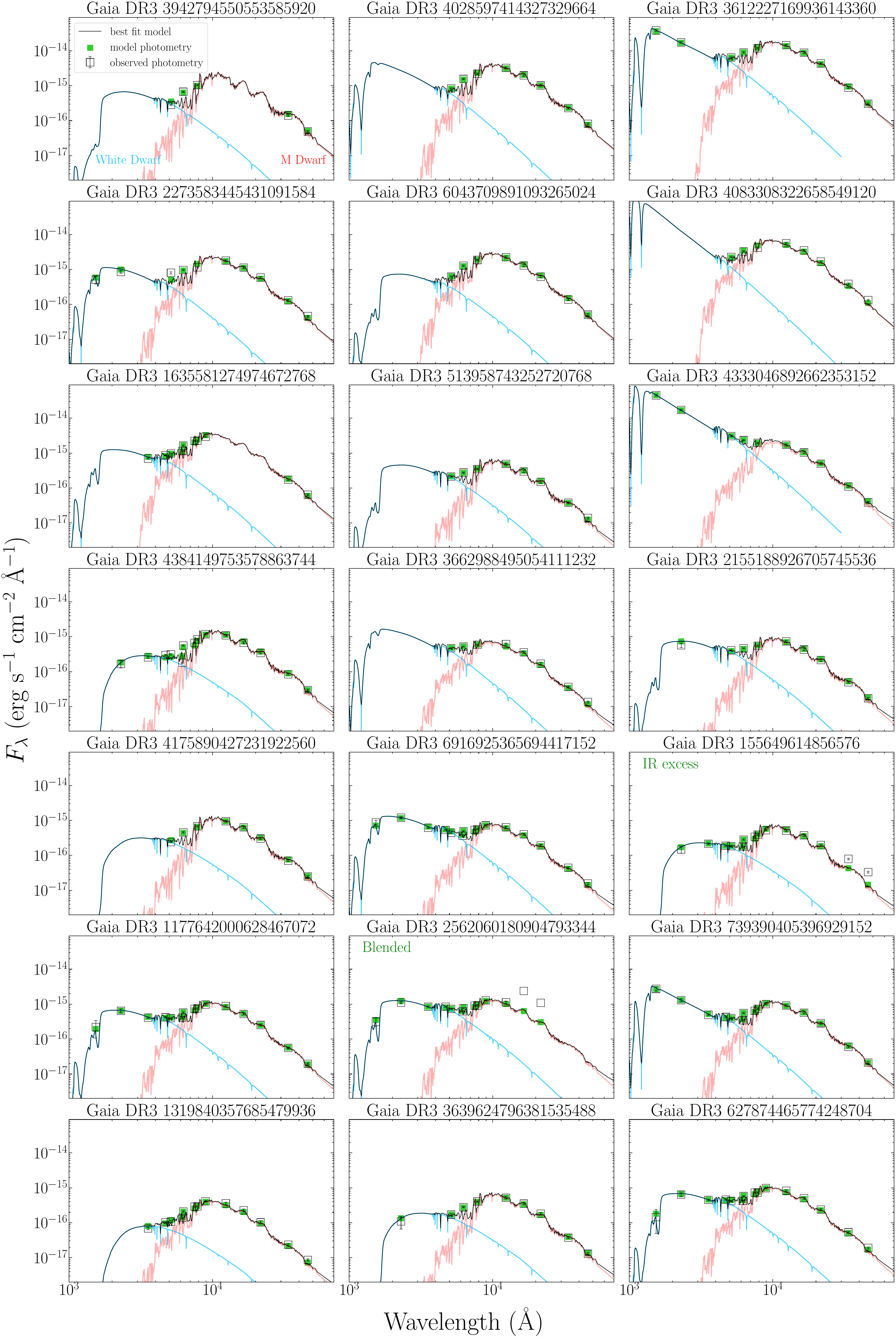}
\caption{
Spectral energy distributions for all PCEBs in our sample.
We show the results of two-component WD + MS SED fitting. 
The observed photometry (black squares) is compared to mock photometry from the best-fit two-component spectrum (green squares).
The best-fit white dwarf (blue) plus MS (red) spectral model (combined in the black line) is also shown, derived using MCMC. 
}
\label{fig:example_SEDfit}
\end{figure}

\begin{figure}
\centering
\includegraphics[width=0.85\columnwidth]{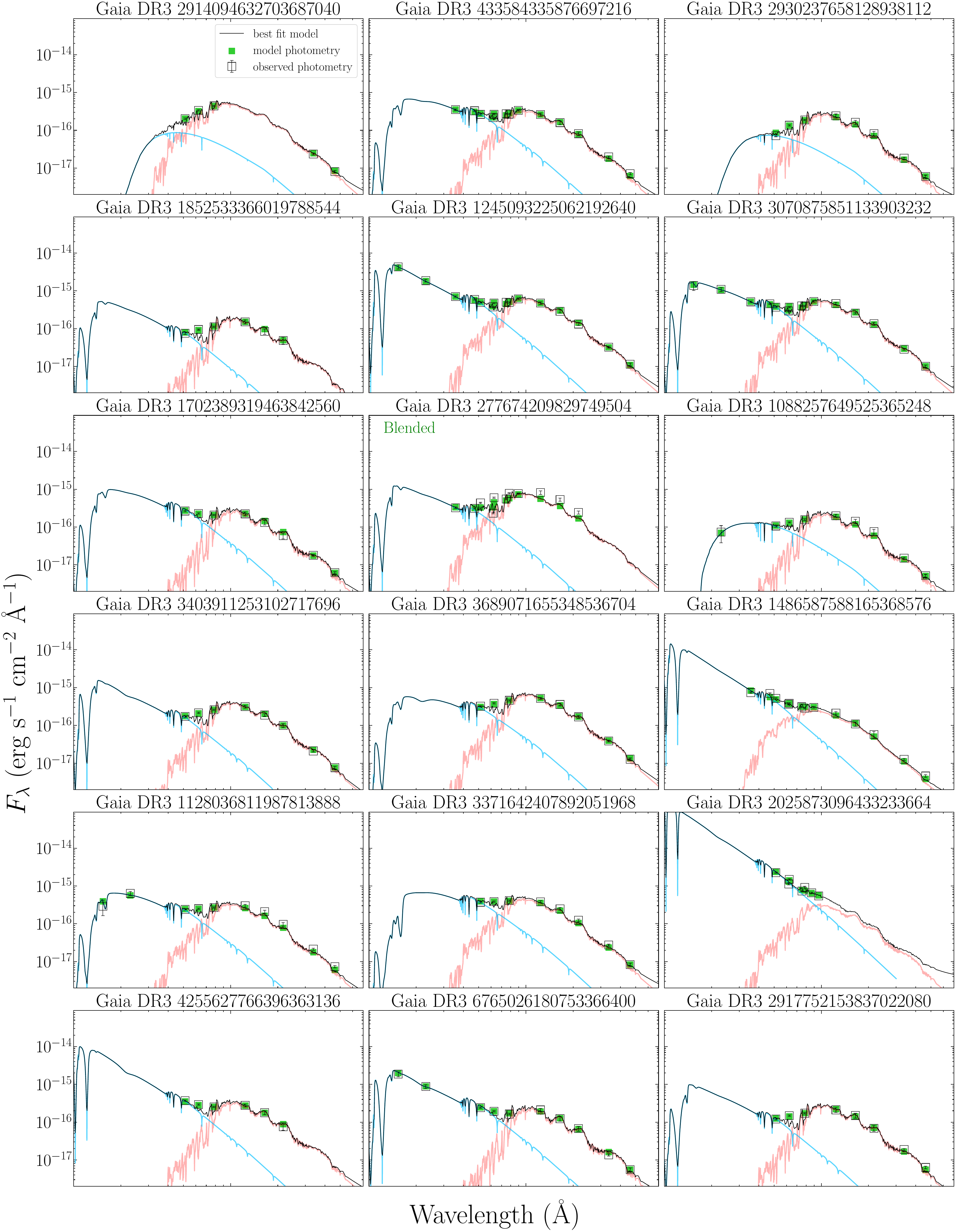}
\caption{
{\bf Continuation of Figure \ref{fig:example_SEDfit}}
}
\label{fig:all_SEDs_2}
\end{figure}
\begin{deluxetable*}{lccccccccccc}
    \centering
    \tablecaption{Stellar parameters from two-component SED fitting for the main-sequence (MS) and white dwarf (WD) in each binary.}
    \tablehead{
    \colhead{{\it Gaia} DR3 ID} &
    \colhead{$\varpi$} &
    \colhead{$P_{\rm orb}$} &
    \colhead{$T_{\rm eff,MS}$} &
    \colhead{$R_{\rm MS}$} &
    \colhead{$M_{\rm MS}$} &
    \colhead{$T_{\rm eff,WD}$} &
    \colhead{$R_{\rm WD}$} &
    \colhead{$M_{\rm WD}$} &
    \colhead{$E(B-V)$} &
    \colhead{$t_{\rm cool}$} &
    \colhead{Ref.} 
    \\
    \colhead{} &
    \colhead{[mas]} &
    \colhead{[hr]} &
    \colhead{[K]} &
    \colhead{[${\rm R_\odot}$]} &
    \colhead{[${\rm M_\odot}$]} &
    \colhead{[K]} &
    \colhead{[$10^{-3}~{\rm R_\odot}$]} &
    \colhead{[${\rm M_\odot}$]} &
    \colhead{[mag]} &
    \colhead{[Gyr]} &
    \colhead{} 
    }
    \startdata
3612227169936143360 & 19.92 & 3.62 & $3097^{+22}_{-20}$ & $0.43^{+0.00}_{-0.01}$ & $0.451^{+0.005}_{-0.007}$ & $13867^{+295}_{-212}$ & $12.98^{+0.31}_{-0.66}$ & $0.626^{+0.020}_{-0.044}$ & 0.0149 & $0.29^{+0.02}_{-0.01}$ & (a) \\
2273583445431091584 & 17.71 & 3.43 & $3020^{+13}_{-33}$ & $0.18^{+0.00}_{-0.00}$ & $0.159^{+0.005}_{-0.002}$ & $11933^{+125}_{-181}$ & $4.23^{+0.03}_{-0.02}$ & $1.298^{+0.002}_{-0.001}$ & 0.0080 & $1.43^{+0.06}_{-0.08}$ & -- \\
4333046892662353152 & 11.04 & 9.15 & $3033^{+35}_{-38}$ & $0.27^{+0.01}_{-0.01}$ & $0.263^{+0.009}_{-0.007}$ & $17023^{+265}_{-246}$ & $16.00^{+0.30}_{-0.37}$ & $0.477^{+0.011}_{-0.014}$ & 0.0106 & $0.11^{+0.00}_{-0.00}$ & (b) \\
4384149753578863744 & 10.80 & 2.31 & $2795^{+6}_{-26}$ & $0.27^{+0.01}_{-0.00}$ & $0.261^{+0.008}_{-0.003}$ & $7805^{+881}_{-126}$ & $11.56^{+0.12}_{-1.74}$ & $0.724^{+0.008}_{-0.133}$ & 0.0120 & $1.78^{+0.42}_{-0.20}$ & (c) \\
2155188926705745536 & 10.36 & 21.20 & $2846^{+31}_{-29}$ & $0.21^{+0.00}_{-0.01}$ & $0.196^{+0.005}_{-0.006}$ & $10119^{+211}_{-87}$ & $10.09^{+0.24}_{-0.26}$ & $0.834^{+0.020}_{-0.022}$ & 0.0096 & $1.17^{+0.01}_{-0.06}$ & -- \\
6916925365694417152 & 8.56 & 13.87 & $3005^{+12}_{-12}$ & $0.21^{+0.00}_{-0.00}$ & $0.198^{+0.002}_{-0.002}$ & $11241^{+86}_{-108}$ & $11.38^{+0.08}_{-0.06}$ & $0.737^{+0.006}_{-0.004}$ & 0.0076 & $0.64^{+0.03}_{-0.14}$ & (d) \\
155649614856576 & 8.30 & 3.23 & $2457^{+10}_{-9}$ & $0.35^{+0.00}_{-0.00}$ & $0.359^{+0.005}_{-0.004}$ & $8015^{+82}_{-110}$ & $14.42^{+0.10}_{-0.08}$ & $0.545^{+0.005}_{-0.004}$ & 0.0057 & $1.00^{+0.04}_{-0.00}$ & (e) \\
1177642000628467072 & 8.19 & 5.74 & $3124^{+11}_{-10}$ & $0.25^{+0.00}_{-0.00}$ & $0.240^{+0.003}_{-0.003}$ & $10489^{+92}_{-86}$ & $10.53^{+0.07}_{-0.08}$ & $0.798^{+0.005}_{-0.006}$ & 0.0096 & $0.84^{+0.01}_{-0.01}$ & (f) \\
2562060180904793344 & 7.82 & 5.03 & $3055^{+49}_{-79}$ & $0.30^{+0.02}_{-0.01}$ & $0.293^{+0.028}_{-0.017}$ & $10005^{+117}_{-655}$ & $17.48^{+2.85}_{-0.40}$ & $0.424^{+0.072}_{-0.013}$ & 0.0086 & $0.56^{+0.00}_{-0.07}$ & -- \\
739390405396929152 & 7.47 & 3.10 & $3011^{+13}_{-13}$ & $0.30^{+0.00}_{-0.00}$ & $0.305^{+0.005}_{-0.005}$ & $13951^{+174}_{-239}$ & $8.78^{+0.17}_{-0.14}$ & $0.943^{+0.014}_{-0.013}$ & 0.0074 & $0.70^{+0.00}_{-0.07}$ & (g) \\
3639624796381535488 & 7.04 & 13.70 & $2892^{+24}_{-169}$ & $0.26^{+0.01}_{-0.00}$ & $0.255^{+0.007}_{-0.005}$ & $8071^{+230}_{-131}$ & $13.92^{+6.32}_{-0.21}$ & $0.571^{+0.218}_{-0.011}$ & 0.0287 & $1.00^{+0.11}_{-0.00}$ & -- \\
627874465774248704 & 6.98 & 11.55 & $3172^{+13}_{-12}$ & $0.27^{+0.00}_{-0.00}$ & $0.266^{+0.003}_{-0.004}$ & $10178^{+120}_{-208}$ & $13.54^{+0.30}_{-0.10}$ & $0.592^{+0.017}_{-0.005}$ & 0.0122 & $0.56^{+0.06}_{-0.00}$ & -- \\
1245093225062192640 & 5.99 & 5.96 & $3140^{+18}_{-17}$ & $0.26^{+0.00}_{-0.00}$ & $0.245^{+0.004}_{-0.004}$ & $14354^{+271}_{-262}$ & $13.07^{+0.33}_{-0.32}$ & $0.620^{+0.020}_{-0.020}$ & 0.0124 & $0.28^{+0.05}_{-0.02}$ & (h) \\
3070875851133903232 & 5.91 & 10.05 & $3050^{+18}_{-18}$ & $0.26^{+0.00}_{-0.00}$ & $0.246^{+0.005}_{-0.004}$ & $12276^{+197}_{-148}$ & $13.48^{+0.16}_{-0.24}$ & $0.596^{+0.009}_{-0.015}$ & 0.0135 & $0.32^{+0.01}_{-0.01}$ & -- \\
1088257649525365248 & 5.65 & 2.37 & $2838^{+31}_{-36}$ & $0.20^{+0.00}_{-0.00}$ & $0.178^{+0.005}_{-0.005}$ & $7243^{+170}_{-104}$ & $17.49^{+0.21}_{-0.28}$ & $0.424^{+0.007}_{-0.009}$ & 0.0097 & $1.26^{+0.14}_{-0.00}$ & -- \\
1128036811987813888 & 5.40 & 33.17 & $3097^{+17}_{-17}$ & $0.21^{+0.00}_{-0.00}$ & $0.197^{+0.003}_{-0.003}$ & $11275^{+138}_{-114}$ & $12.54^{+0.11}_{-0.11}$ & $0.654^{+0.008}_{-0.008}$ & 0.0073 & $0.50^{+0.02}_{-0.03}$ & -- \\
6765026180753366400 & 5.16 & 2.01 & $2723^{+33}_{-15}$ & $0.24^{+0.00}_{-0.01}$ & $0.228^{+0.005}_{-0.008}$ & $13880^{+202}_{-210}$ & $12.26^{+0.23}_{-0.23}$ & $0.675^{+0.016}_{-0.016}$ & 0.0416 & $0.29^{+0.03}_{-0.01}$ & -- \\
3942794550553585920$^{*}$ & 22.17 & 13.34 & $2640^{+11}_{-10}$ & $0.18^{+0.00}_{-0.00}$ & $0.161^{+0.002}_{-0.002}$ & $10079^{+176}_{-51}$ & $4.29^{+0.04}_{-0.05}$ & -- & 0.0042 & $2.41^{+0.11}_{-0.03}$ & (d) \\
4028597414327329664$^{*}$ & 21.79 & 2.99 & $2912^{+17}_{-15}$ & $0.21^{+0.00}_{-0.00}$ & $0.191^{+0.003}_{-0.003}$ & $14557^{+508}_{-400}$ & $3.81^{+0.06}_{-0.02}$ & -- & 0.0074 & $0.76^{+0.07}_{-0.06}$ & (h) \\
6043709891093265024$^{*}$ & 16.50 & 11.60 & $3090^{+19}_{-23}$ & $0.21^{+0.00}_{-0.00}$ & $0.190^{+0.004}_{-0.004}$ & $10586^{+1486}_{-1185}$ & $5.31^{+0.95}_{-0.89}$ & -- & 0.0047 & $2.11^{+0.74}_{-0.73}$ & -- \\
4083308322658549120$^{*}$ & 14.58 & 4.96 & $3203^{+5}_{-2}$ & $0.34^{+0.00}_{-0.00}$ & $0.346^{+0.003}_{-0.004}$ & $26041^{+5070}_{-7423}$ & $5.11^{+1.88}_{-0.94}$ & -- & 0.0102 & $0.13^{+0.07}_{-0.25}$ & -- \\
1635581274974672768$^{*}$ & 13.44 & 2.79 & $3137^{+52}_{-68}$ & $0.27^{+0.02}_{-0.01}$ & $0.267^{+0.030}_{-0.013}$ & $10734^{+3835}_{-2636}$ & $8.14^{+6.38}_{-2.43}$ & -- & 0.0083 & $2.03^{+1.27}_{-1.85}$ & -- \\
513958743252720768$^{*}$ & 11.09 & 2.38 & $2797^{+34}_{-40}$ & $0.17^{+0.00}_{-0.00}$ & $0.148^{+0.005}_{-0.005}$ & $10244^{+2288}_{-2776}$ & $6.80^{+5.51}_{-1.54}$ & -- & 0.0092 & $2.31^{+1.13}_{-2.17}$ & -- \\
3662988495054111232$^{*}$ & 10.77 & 2.93 & $3235^{+27}_{-24}$ & $0.14^{+0.00}_{-0.00}$ & $0.120^{+0.002}_{-0.002}$ & $12477^{+498}_{-417}$ & $7.27^{+0.11}_{-0.12}$ & -- & 0.0051 & $1.20^{+0.15}_{-0.17}$ & (i) \\
4175890427231922560$^{*}$ & 9.83 & 2.68 & $2784^{+14}_{-18}$ & $0.27^{+0.00}_{-0.00}$ & $0.266^{+0.005}_{-0.004}$ & $8053^{+275}_{-105}$ & $13.05^{+0.13}_{-0.31}$ & -- & 0.0320 & $1.14^{+0.10}_{-0.11}$ & -- \\
1319840357685479936$^{*}$ & 7.07 & 3.89 & $3162^{+29}_{-26}$ & $0.18^{+0.00}_{-0.00}$ & $0.157^{+0.004}_{-0.004}$ & $7533^{+738}_{-345}$ & $10.26^{+0.37}_{-0.92}$ & -- & 0.0234 & $2.51^{+0.51}_{-1.03}$ & -- \\
2914094632703687040$^{*}$ & 6.66 & 3.67 & $3342^{+30}_{-41}$ & $0.19^{+0.00}_{-0.00}$ & $0.168^{+0.005}_{-0.004}$ & $6092^{+459}_{-282}$ & $19.36^{+0.53}_{-0.67}$ & -- & 0.0067 & $2.00^{+0.41}_{-0.24}$ & -- \\
433584335876697216$^{*}$ & 6.10 & 8.63 & $3114^{+40}_{-53}$ & $0.19^{+0.01}_{-0.01}$ & $0.169^{+0.006}_{-0.006}$ & $11321^{+1177}_{-1114}$ & $12.51^{+1.86}_{-1.43}$ & -- & 0.0360 & $0.50^{+0.13}_{-0.19}$ & -- \\
2930237658128938112$^{*}$ & 6.08 & 2.60 & $2905^{+43}_{-39}$ & $0.19^{+0.01}_{-0.01}$ & $0.175^{+0.007}_{-0.007}$ & $5905^{+142}_{-180}$ & $19.82^{+0.33}_{-0.25}$ & -- & 0.0135 & $2.24^{+0.02}_{-0.00}$ & -- \\
1852533366019788544$^{*}$ & 6.00 & 25.25 & $2831^{+74}_{-447}$ & $0.17^{+0.08}_{-0.01}$ & $0.151^{+0.094}_{-0.014}$ & $13829^{+3671}_{-7217}$ & $5.37^{+12.78}_{-0.97}$ & -- & 0.0121 & $0.87^{+0.41}_{-4.62}$ & -- \\
1702389319463842560$^{*}$ & 5.80 & 4.84 & $2774^{+23}_{-23}$ & $0.22^{+0.00}_{-0.00}$ & $0.206^{+0.004}_{-0.004}$ & $11748^{+323}_{-243}$ & $12.74^{+0.18}_{-0.18}$ & -- & 0.0155 & $0.48^{+0.07}_{-0.03}$ & (j) \\
277674209829749504$^{*}$ & 5.78 & 36.14 & $3229^{+18}_{-23}$ & $0.28^{+0.00}_{-0.00}$ & $0.274^{+0.005}_{-0.004}$ & $12693^{+1089}_{-1654}$ & $10.31^{+2.21}_{-0.94}$ & -- & 0.0175 & $0.71^{+0.15}_{-0.23}$ & -- \\
3403911253102717696$^{*}$ & 5.59 & 5.42 & $2987^{+48}_{-70}$ & $0.25^{+0.01}_{-0.01}$ & $0.234^{+0.013}_{-0.010}$ & $14396^{+4511}_{-4057}$ & $8.58^{+4.33}_{-1.86}$ & -- & 0.0479 & $0.70^{+0.45}_{-1.04}$ & (d) \\
3689071655348536704$^{*}$ & 5.51 & 3.99 & $3026^{+35}_{-43}$ & $0.32^{+0.01}_{-0.01}$ & $0.326^{+0.012}_{-0.010}$ & $11234^{+1631}_{-1681}$ & $15.99^{+4.56}_{-1.86}$ & -- & 0.1001 & $0.40^{+0.12}_{-0.25}$ & -- \\
1486587588165368576$^{*}$ & 5.44 & 3.02 & $3594^{+22}_{-17}$ & $0.14^{+0.00}_{-0.00}$ & $0.117^{+0.002}_{-0.002}$ & $18360^{+293}_{-232}$ & $11.17^{+0.13}_{-0.14}$ & -- & 0.0104 & $0.15^{+0.01}_{-0.01}$ & (e) \\
3371642407892051968$^{*}$ & 5.39 & 6.22 & $3168^{+22}_{-25}$ & $0.24^{+0.00}_{-0.00}$ & $0.226^{+0.004}_{-0.004}$ & $10006^{+90}_{-95}$ & $19.08^{+0.15}_{-0.13}$ & -- & 0.0172 & $0.56^{+0.00}_{-0.01}$ & (d) \\
2025873096433233664$^{*}$ & 5.22 & 3.58 & $3159^{+75}_{-96}$ & $0.21^{+0.01}_{-0.01}$ & $0.195^{+0.013}_{-0.009}$ & $21542^{+2242}_{-2988}$ & $25.02^{+3.65}_{-1.91}$ & -- & 0.0283 & $0.04^{+0.02}_{-0.04}$ & (j) \\
4255627766396363136$^{*}$ & 5.18 & 8.73 & $2877^{+107}_{-91}$ & $0.26^{+0.02}_{-0.02}$ & $0.253^{+0.019}_{-0.021}$ & $20444^{+8133}_{-7714}$ & $11.20^{+5.12}_{-3.21}$ & -- & 0.0760 & $0.11^{+0.09}_{-0.36}$ & -- \\
2917752153837022080$^{*}$ & 5.15 & 4.64 & $2864^{+71}_{-92}$ & $0.24^{+0.02}_{-0.01}$ & $0.224^{+0.022}_{-0.014}$ & $13710^{+6589}_{-5144}$ & $7.88^{+7.61}_{-2.17}$ & -- & 0.0110 & $0.89^{+0.59}_{-2.59}$ & -- \\
    \enddata
    \tablecomments{Columns: (1) Gaia DR3 source ID; (2) parallax in milliarcseconds; (3) orbital period in hours; (4-6) effective temperature, radius, and mass of the WD; (7-9) effective temperature, radius, and mass of the MS; (10) extinction from the \citet{Edenhofer24} dust map; (11) WD cooling age; (12) reference. The references are (a) \citet{Marsh00}, (b) \citet{Bravo25}, (c) \citet{Brown22}, (d) \citet{Brown23}, (e) \citet{Silvestri07}, (f) \citet{Parsons15}, (g) \citet{RebassaMansergas13}, (h) \citet{RebassaMansergas10}, (i) \citet{Farihi05}, and (j) \citet{Kosakowski22}. `*' sources do not have GALEX UV photometry available. A machine-readable version of the table is available at \url{https://github.com/cheyanneshariat/pcebs}.\label{tab:WDMS_params}}
\end{deluxetable*}

\clearpage
\bibliography{references}

\end{document}